\documentclass[referee]{mn2e}
\pdfoutput=1
\usepackage{graphicx}

\def\T1{\ {$T_1$}\ }
\def\MT1{\ {$M_{T_1}$}\ }
\def\ct1{\ {$(C-T_1)$}\ }
\def\CT10{\ {$(C-T_1)_0$}\ }

\def\VI0{\ {$(V-I)_0$}\ }

\def\2cd{\ {two-color diagram}\ }

\def\sf{\ {specific frequency}\ }
\def\ell{\ {elliptical}\ }

\def\gtsim{\ {\raise-0.5ex\hbox{$\buildrel>\over\sim$}}\ }
\def\ltsim{\ {\raise-0.5ex\hbox{$\buildrel<\over\sim$}}\ }

\begin{document}

\title[Intermediate-age LMC clusters]{Gemini/GMOS photometry of intermediate-age star clusters
in the Large Magellanic Cloud}

\author[Piatti et al.]{Andr\'es E. Piatti$^{1,2}$, Stefan C. Keller$^{3}$, A. Dougal Mackey$^{3}$,
\newauthor and Gary S. Da Costa$^{3}$\\
$^1$ Observatorio Astron\'omico, Universidad Nacional de C\'ordoba, Laprida 854, 5000 C\'ordoba, Argentina\\
$^2$Consejo Nacional de Investigaciones Cient\'{\i}ficas y T\'ecnicas, Av. Rivadavia 1917, C1033AAJ, Buenos Aires, Argentina \\
$^3$ Research School of Astronomy and Astrophysics, Australian National University, Canberra, Australia}

\maketitle

\begin{abstract}
We present Gemini South GMOS $g,i$ photometry 
of 14 intermediate-age Large Magellanic Cloud (LMC) star clusters, namely: NGC\,2155, 2161, 2162, 
2173, 2203, 2209, 2213, 2231, 2249, Hodge\,6, SL\,244, 505, 674, and 769, as part of a continuing
 project to investigate the extended Main Sequence Turnoff (EMSTO) phenomenon. Extensive artificial star tests
were made over the observed field of view. These tests reveal the observed behaviour of photometric errors 
with magnitude and crowding. 
The cluster stellar density radial profiles were traced from star counts over the extent of the observed field.
We adopt cluster radii and build colour-magnitude diagrams (CMDs) with
cluster features clearly identified.
We used the cluster ($g,g-i$) CMDs to estimate ages from the matching of theoretical 
isochrones. The studied LMC clusters are confirmed to be intermediate-age
clusters, which range in age 9.10 $<$ log($t$) $<$ 9.60. NGC\,2162 and NGC\,2249 look like new 
EMSTO candidates, in addition to NGC\,2209, on the basis of having dual red clumps.
\end{abstract}


\section{Introduction}
The massive stellar clusters in the Large Magellanic Cloud (LMC) possess a wide range of ages
(10$^{6}$ to 10$^{10}$ yr) though as well as a well-established dearth of clusters with ages between 
$\sim$ 3-4$\times$10$^{9}$ and 10$^{10}$ yr (see, e.g., Da Costa 1991, Baumgardt et al. 2013). 
This makes the LMC an important testing ground for theories of how massive star clusters form
 and evolve both dynamically and possibly chemically. Because of the proximity of the LMC, massive 
clusters of intermediate-age (i.e., 1-3 Gyr) are readily accessible for detailed study.
The cluster NGC\,1846 (M $\approx$ 10$^{5}$ M$_{\odot}$; age $\approx$ 1.8 Gyr; 
[Fe/H] $\approx$  -0.4) was found by Mackey \& Broby Nielsen \shortcite{mbn07} to possesses a colour-magnitude 
diagram (CMD) exhibiting two distinct main-sequence turn-offs (MSTOs). The width of the red giant branch 
is small, which indicates there is no significant internal spread in [Fe/H]. The cluster is consistent
 with the presence of two stellar populations with ages 300 Myr apart \cite{metal08}.

Searches of the HST/ACS archive \cite{metal08,metal09} revealed another 10 LMC clusters of ages between
 1 -- 2.5 Gyr that exhibit unusual MSTOs. The MSTO region of these clusters may
be bifurcated or much more extended than can be accounted for by photometric errors. Binary stars 
seem to contribute at different levels to the broadness around the MSTO 
\cite{metal09,getal09,yetal11}. Similarly 
stellar rotation appears  in some works as an unlikely solution \cite{gietal09}, or performs equally well
like age spread \cite{letal14}, or is preferred \cite{bm09}. Milone et al. report
that 11 of 16 (i.e. ($70 \pm 25$)\%) intermediate-age clusters possess this phenomenon.

Keller, Mackey \& Da Costa \shortcite{ketal11} have simulated stellar populations with a range of 
luminosities and star formation histories. These simulations show that a cluster with bimodal star 
formation history featuring a 200 Myr hiatus would be undetectable
to existing ACS photometry of LMC clusters once the age of the cluster exceeds 2.3 Gyr.
This is a consequence of the fact that in increasingly older clusters, the difference in age between the
 constituent stellar populations represents a diminishing fraction of the cluster age, hence the multiple
 populations
become increasingly harder to resolve photometrically. The fact that the multiple MSTO phenomenon is only
detected in globular clusters (GCs) younger than 2.5 Gyr is therefore not unexpected. Indeed, it is plausible 
that the production of
multiple populations is an evolutionary phase for the majority of massive clusters, although it is not seen
in some massive younger clusters \cite{bsv13,cetal14}.
The currently known multiple MSTO clusters possess another outstanding feature. The extended Main Sequence 
Turnoff (EMSTO) clusters show a tendency to have larger core radii than non-EMSTO clusters \cite{ketal11}.


In order to investigate the frequency of occurrence of the multiple MSTO phenomenon in the LMC cluster
population, we have built an age and luminosity limited sample with which to examine
if the formation of multiple populations is a general phase of cluster evolution, one possibly related to
 the puzzling multiple populations seen in ancient Galactic GCs (see for example, Carretta et al. 2010). 
In this paper we present a photometric dataset for a number
 of clusters in the sample for which high-quality CMDs were previously unavailable. 
Although we defer a detailed analysis on the presence or absence of the extended MSTO phenomenon in these 
clusters to a forthcoming paper (Keller et al. 2014, in prep.), we investigate the status of each cluster 
 on the basis of matching single stellar population (SSP) isochrones to their respective 
CMDs. We thus provide estimates of the mean cluster ages, which will assist in the analysis of the role of 
cluster core radii, the degree of MSTO broadness, dynamical status of the clusters, etc. Results from this
 program for the LMC cluster NGC\,2209 have been published in Keller et al. \shortcite{ketal12}.
 In Section 3 we present the derived
 cluster photometry, followed by derivation of the fundamental cluster parameters of age and metallicity, 
and spatial extent in Section 4.


\section{Observations and Data reduction}

We obtained images of 14 candidate LMC intermediate-age clusters with the Gemini South telescope 
and the GMOS-S instrument through $g$ and $i$ filters.  In imaging mode GMOS-S has a field-of-view
 of approximately 5.5$\arcmin$$\times$ 5.5$\arcmin$ at a scale of 0.146 arcsec per (2x2 binned) pixel.  
The detector is a 3$\times$1 mosaic of 2K$\times$4K EEV CCDs.  Observations were executed in 
queue mode (under programs GS-2011A-Q-43, GS-2012A-Q-15 and GS-2013A-Q-17) which enabled the 
data to be obtained in excellent seeing (0.35" to 0.78" FWHM) and under photometric conditions.  
The log of observations is presented in Table 1, where the main 
astrometric, photometric and observational information is summarized.
Several images were taken in each filter ($g,i$), as we judged that the dynamic
range and accuracy required to suit our science goals could be met most efficiently this way.
Most of the selected fields have shorter and longer exposure times to provide coverage of 
bright cluster red giant branch 
stars as well as stars at least two magnitudes below the MSTO in order to search for the presence of 
the extended MSTO phenomenon.

The data reduction followed the procedures documented in the Gemini Observatory 
webpage\footnote{http://www.gemini.edu}
and utilized the {\sc gemini/gmos} package in IRAF\footnote{IRAF is distributed by the National 
Optical Astronomy Observatories, which is operated by the Association of 
Universities for Research in Astronomy, Inc., under contract with the National 
Science Foundation.}. We performed overscan, trimming, bias subtraction, flattened all data images, etc., 
once the
calibration frames (zeros and flats) were properly combined.

Observations of photometric standard stars were included in the baseline calibrations for GMOS. 
The standard stars were chosen from the standard star catalog calibrated directly
in the SDSS system (Smith et al. 2014, http://www-star.fnal.gov). For program GS-2011A-Q-43, the
standard fields 160100-600000, E5$\_$b, E3$\_$a. PG1633+099, and LSE\,259 were observed during the same nights
as for NGC\,2155, 2161, 2162, 2173, 2203, 2209, 2213, and 2231. NGC\,2231 was observed during 
two different nights, so that we used its photometry for additional controls. The calibrated photometry for
NGC\,2173 and 2209 have been previously published in Keller et al. \shortcite{ketal12}. For program
GS-2013A-Q-17, the standard fields 060000-300000 and 160100-600000 were observed during the same night
as for clusters SL\,244, 505, 674, and 769. Images of standard star fields were not observed as a 
regular-based instrument monitoring (program objects NGC\,2249 and Hodge\,6). However, we have paid 
particular care when using the calibrations derived (see also, http://www.gemini.edu/node/10625?q=node/10445) 
assuming that the atmospheric extinction was close to the median value for Cerro Pachon. 

Independent magnitude measures of standard stars
were derived per filter using
the {\sc apphot} task within IRAF, in order to secure the transformation
from the instrumental to the SDSS $gi$ standard system.  Standard stars were
distributed over an area similar to that of the GMOS array, so that we measured magnitudes of standard stars in 
each of the three chips.  The relationships between
instrumental and standard magnitudes were obtained by fitting
the following equations:

\begin{equation}
g = g_1 + g_{std} + g_2\times X_g + g_3\times (g-i)_{std}
\end{equation}

\begin{equation}
i = i_1 + i_{std} + i_2\times X_i + i_3\times (g-i)_{std}
\end{equation}

\noindent where $g_j$, and $i_j$ (j=1,3) are the fitted coefficients, and
$X$ represents the effective airmass. We solved the transformation equations with the {\sc fitparams}
task in IRAF.
The root-mean square (rms) errors from
the transformation to the standard system were 0.011 mag for $g$ and 0.013 for $i$, respectively, 
indicating excellent photometric quality.

The stellar photometry was performed using the star-finding and point-spread-function (PSF) fitting 
routines in the {\sc daophot/allstar} suite of programs \cite{setal90}. For each frame, a quadratically varying 
PSF was derived by fitting $\sim$ 60 stars, once the neighbours were eliminated using a preliminary PSF
derived from the brightest, least contaminated 20-30 stars. Both groups of PSF 
stars were interactively selected. We then used the {\sc allstar} program to apply the resulting PSF to the 
identified stellar objects and to create a subtracted image which was used to find and measure magnitudes of 
additional fainter stars. This procedure was repeated three times for each frame. Finally, 
we computed aperture corrections from the comparison of PSF and aperture magnitudes by using the 
neighbour-subtracted PSF star sample. After deriving the photometry for all detected objects in
each filter, a cut was made on the basis of the parameters
returned by {\sc daophot}. Only objects with $\chi$ $<$2, photometric error less than 2$\sigma$ above 
the mean error at a given magnitude, and $|$SHARP$|$ $<$ 0.5 were kept in each filter
(typically discarding about 10\% of the objects), and then the
remaining objects in the $g$ and $i$ lists were matched with a
tolerance of 1 pixel and raw photometry obtained. 

We combined all the independent instrumental magnitudes using the stand-alone {\sc daomatch} and 
{\sc daomaster} programs\footnote{Program kindly provided by P.B. Stetson}. As a result, we produced 
one dataset per cluster containing
the $x$ and $y$ coordinates for each star, and different ($g$,$g-i$) pairs according to
the number of frames obtained per filter. We did not combine ($g$,$i$) shorter with longer exposures,
 but treat them 
separately. The gathered photometric information were standardized using equations (1) to (2). 
We finally  averaged 
standard magnitudes and colours of stars measured several times. The resulting standardized photometric 
tables
consist of a running number per star, equatorial coordinates, the averaged $g$ 
magnitudes and $g-i$ colours, their respective rms errors $\sigma(g)$ and 
$\sigma(g-i)$, and the number of observations per star. We adopted 
the photometric errors provided by {\sc allstar} for stars 
with only one measure. Tables 2 to 15 provide this information for  NGC\,2155, 2161, 2162, 2173, 
2203, 2209,
2213, 2231, 2249, Hodge\,6, SL\,244, 505, 674, and 769, respectively. Only a portion 
of Table 2 is shown here for guidance regarding their form and content. The whole content of 
Tables 2-15 is 
available in the online version of the journal on Oxford journals, at 
http://access.oxfordjournals.org.

We first examined the quality of our photometry in order to evaluate the influence of the photometric 
errors, crowding effects and the detection limit on the cluster fiducial characteristics in the CMDs. 
To do this, we performed artificial star tests on a long exposure image per filter and per cluster 
to derive the completeness
level at different magnitudes. We used the stand-alone {\sc addstar} program in the {\sc daophot}
package \cite{setal90} to add synthetic stars, generated bearing in mind the colour and magnitude distributions 
of the stars in the CMDs (mainly along the main sequence and the red giant branch), as well as the radial stellar 
density profiles 
of the cluster fields. We added a number
of stars equivalent to $\sim$ 5$\%$ of the measured stars in order to avoid
significantly more crowding synthetic images than in the original images. On the other hand, to avoid small
 number statistics in the artificial-star 
analysis, 
we created five different images for each one used in the artificial star tests. 
Utilizing the tabulated gains for the GMOS devices we  were
 able to simulate the Poisson noise in each stellar image. 


We then repeated the same steps to obtain the photometry of the synthetic images as described above, i.e., 
performing three passes with the {\sc daophot/allstar} routines, making a cut on the basis of the parameters
returned by {\sc daophot}, etc. 
The errors and star-finding efficiency was estimated by comparing the output 
and the input data for these stars - within the respective magnitude and colour bins - 
using the {\sc daomatch} and {\sc daomaster} tasks.
In Fig. 1 we show the resultant completeness fractions as a function of magnitude for NGC\,2173, which is
a representative cluster in our sample when considering simultaneously the largest extent, crowding and number of measured stars.  
Fig. 1 shows that the 50$\%$ completeness level is reached at $g,i$ $\sim$ 23.5-25.0, 
depending on the distance from the cluster centre. On the other hand, by using the theoretical 
isochrones of Marigo et al. \shortcite{metal08} and the LMC distance modulus $(m-M)_o$ = 18.49 
\cite{detal14},
we concluded that
the MSTO of star clusters with ages between 1-3 Gyr is located at $g$ $\sim$ 20-21 mag.
Thus, we conclude that our photometry is able to reach the 50\% completeness level 2-3 magnitudes below
the MSTO for the innermost cluster regions (distance to the cluster centre $\le$ $r_{HM}$/4 
where $r_{HM}$ is a measure of the cluster size defined in the next section).
The behaviour of $\sigma$($g$), and $\sigma$($g-i$) is represented by error bars in the CMDs shown in
Figs. 2-15.

\section{Analysis of the Colour-magnitude diagrams}

In Fig. 2-15 we show the Colour-Magnitude Diagrams (CMDs) of stars in the field of 
NGC\,2155, 2161, 2162, 2173, 2203, 2209, 2213, 2231, 2249, Hodge\,6, SL\,244, 505, 674, and 769, respectively.
The most obvious traits in each cluster CMD are the long cluster Main Sequence (MS)
which extends over a range of approximately 4-5 mag in $g$, the cluster red clump (RC) and 
red giant branch (RGB). In some cases, a populous cluster sub-giant branch (SGB) is also visible (e.g., NGC\,2155; Fig. 2).
The RC is not tilted in any of the studied clusters - except SL\,244 - so that  differential 
reddening can be assumed to be negligible along the lines of sight. However, NGC\,2162, 2209, and 2249 have 
clear dual RCs, a feature seen in star clusters exhibiting the EMSTO phenomenon \cite{metal09}. 
Keller, Mackey \& Da Costa \shortcite{ketal12} discuss the parameters of the EMSTO evident in NGC 2209. The 
consequences of the detection of the EMSTO phenomenon or otherwise in this set of clusters is the focus 
of a separate paper (Keller et al. 2014 in prep.), for which we will use as inputs the mean cluster age estimates derived in the present work.

We determined the cluster geometrical centres in order to obtain circular extracted CMDs where the 
fiducial features of the clusters could be clearly seen. 
The coordinates of the cluster centres and their estimated uncertainties were determined by fitting Gaussian 
distributions to the star counts in the $x$ and $y$ directions for each cluster. The fits of the Gaussians were 
performed using the {\sc ngaussfit} routine in the {\sc stsdas/iraf} package. We adopted a single Gaussian
 and fixed 
the constant 
to the corresponding background levels (i.e. stellar field densities assumed to be uniform) and the linear terms to 
zero. The centre of the Gaussian, its amplitude and its full width at half-maximum acted as variables. 
The number of stars projected along the $x$ and $y$ directions were counted within intervals of 40 pixel wide. In 
addition, we checked that using spatial bins from 20 to 60 pixels does not result in 
significant changes in the derived centres. 
Cluster centres were finally determined with a typical standard deviation of 
$\pm$ 10 pixels  ($\sim$ 0.3 pc) in all cases.

We then constructed the cluster radial profiles based on star counts previously performed 
within boxes of 40 pixels a side distributed 
throughout the whole field of each cluster. The selected size of the box allowed us to sample statistically 
the stellar spatial distribution.
Thus, the number of stars per unit area at a given radius, $r$, can be directly calculated through 
the expression:

\begin{equation}
(n_{r+20} - n_{r-20})/(m_{r+20} - m_{r-20}),
\end{equation}

\noindent where $n_j$ and $m_j$ represent the number of stars and boxes included in a circle of radius $j$, 
respectively. Note that this method does not necessarily require a complete circle of radius $r$ within the
 observed 
field to estimate the mean stellar density at that distance. This is an important consideration since having 
a stellar 
density profile which extends far away from the cluster centre allows us to estimate the background level with 
high precision. This is necessary to derive the cluster radius ($r_{cls}$), defined as
the distance from the cluster centre where the combined cluster plus background stellar 
density profile is no longer readily distinguished from a constant background value  within 1-$\sigma$
of its fluctuation, which typically led to uncertainties of $\sigma$($r_{cls}$) $\approx$ 2 pc.
 It is also helpful to measure the full width at half-maximum of the 
stellar density profile, which plays a significant role - from a stellar content point of view - in the
 construction of the cluster CMDs. 

The resulting density profiles expressed as number of stars per unit area are shown in the upper right
panel of Figs. 2-15. In these figures, we show the region around the centre of each cluster out to $\sim$ 2.7' 
($\approx$ 1100 pixels). 
The background region surrounding each cluster was delimited between the observed field boundaries and the 
cluster
radius from the cluster's centre. The vertical lines represent the radii at the full width at half-maximum
($r_{HM}$) and $r_{cls}$. The $r_{cls}$ values were estimated by eye on the cluster radial profile plots
according to the above definition,
whereas $r_{HM}$ were calulated from the half-maximum of the cluster radial profiles 
 ($\sigma$($r_{HM}$) $\approx$ 0.5 pc).
Notice that these radial scales are not precisely defined, but that small
 changes in their values does not materially affect the appearence of the CMDs.
We then constructed three CMDs covering different circular extractions around each 
cluster as shown in Figs. 2-15 (upper left, bottom left, and bottom right panels). 
The panels in the figures are arranged, from top to bottom and from left to right, in such a way that exhibit 
the 
stellar population variations from the innermost to the outermost regions of the cluster fields. We start with 
the CMD for stars distributed within $r$ $<$ $r_{HM}$, followed by that of the cluster regions delimited by 
$r$ $<$ $r_{cls}$ and finally by the adopted field CMD. The latter was built using a ring centred on the cluster
of area $\pi$$r_{HM}$$^2$ and internal radius $r_{cls}$. We used the CMDs corresponding to the stars within 
$r_{HM}$ as the cluster fiducial sequence references, and used those for $r_{cls}$ to  match theoretical 
isochrones. 
Some field star contamination is unavoidable, though. However, when comparing field and cluster CMDs, the
 differences 
in stellar content become noticeable, as can be 
seen from the upper left and bottom right panels of Figs. 2-15. Particularly, the field CMDs contain 
much fewer stars and are dominated by relatively older MS star
populations, although composed of stars within a wide age range. The CMDs of the cluster 
sample, on the other hand, exhibit distinct RCs characteristic  of  intermediate-age star clusters around 
1-3 Gyr old.

\section{Determination of fundamental cluster parameters}

We computed $E(B-V)$ colour excesses by interpolating the extinction maps of
Burstein \& 
Heiles \shortcite[hereafter BH]{bh82} using a grid of ({\it l},b) values, with steps of $\Delta$({\it l},b) = 
(0.01$\degr$,0.01$\degr$) covering the observed fields. BH maps were obtained from H\,I (21 cm) emission data for
the southern sky. They furnish us with $E(B-V)$ colour excesses which depend on the Galactic coordinates.
We obtained between 80 and 100 
colour excesses per cluster field. Then, we built histograms and 
calculated their centres and full width at half-maxima (FWHMs). Since the FWHMs values turned out to be
 considerably low ($\sim$ 0.03 mag), we 
concluded that the interstellar absorption is uniform across the cluster fields.
Our adopted reddenings are essentially identical to the BH reddenings tabulated for each cluster in 
NED\footnote{http://ned.ipac.caltech.edu/}. Five of our clusters in the periphery of the LMC also have 
reddenings tabulated in NED on the system of Schlafly \& Finkbeiner \shortcite{sf11}, which is a 
recalibration of  the Schlegel, Finkbeiner \& Davis \shortcite{sfd98} reddenings.  For these clusters the 
mean difference between our adopted reddenings and those from Schlafly \& Finkbeiner \shortcite{sf11} is 
0.01 mag with a standard deviation of 0.02 mag. 
Table 16 lists the adopted $E(B-V)$ colour excesses, from which we computed the $E(g-i)$ and $A_{g}$
values  using the $E(g-i)$/$E(B-V)$ = 1.621 and $A_{g}$/$E(B-V)$ = 3.738 ratios given by Cardelli, 
Clayton \& Mathis \shortcite{cetal89}.

As for the cluster distance moduli, Subramanian \& Subramanian 
\shortcite{ss09} find that the average depth for the LMC disc is 3.44 $\pm$ 1.16 kpc, so that 
the difference in apparent distance modulus - clusters could be placed in front of, or behind the LMC -
could be as large as $\Delta$($(m-M)_o$) $\sim$ 0.15 mag, if a value of 50 kpc is adopted for the mean LMC 
distance \cite{detal14}. Since a difference of 0.05 in log($t$) (the difference between two close 
isochrones in the Bressan et al. 2012 models used here) implies a difference of $\sim$ 0.25 mag 
in $g$, we decided to adopt the value of the LMC distance modulus 
$(m-M)_o$ = 18.49 $\pm$ 0.09 reported by de Grijs et al. (2014)  for all the clusters. 
Our simple assumption of adopting a unique value for 
the distance modulus for all the clusters should not dominate the error budget in our final results. In 
fact, when overplotting the Zero-Age Main-Sequence (ZAMS) on the cluster CMDs, previously shifted by 
the corresponding $E(g-i)$  and $(m-M)_o$ = 18.49, excellent matches were generally found.

In order to estimate the cluster ages, it must be taken into account that cluster metallicity 
plays an important role when  matching theoretical isochrones. The 
distinction is mainly evident for the evolved RC and RGB phases. ZAMSs are often less affected by
metallicity effects and can even exhibit imperceptible variations for a specific metallicity range 
within the expected photometric errors. We took advantage of the available theoretical isochrones
computed for the SDSS photometric system to estimate  cluster ages. We used the isochrones 
calculated with core overshooting included by the Padova  group \cite{betal12}.  
When we chose subsets of isochrones for different Z metallicity values to evaluate the 
metallicity effect in the cluster fundamental parameters, we adopted the most frequently 
used value of [Fe/H] = -0.4 dex (Z = 0.006, Z$_{\odot}$ = 0.0152) for the intermediate age LMC 
clusters studied to date \cite[see their Fig. 6]{pg13}.

We then selected a set of isochrones and superimposed them on the cluster CMDs, once they were 
properly shifted by the corresponding LMC distance 
modulus. Notice that by matching different SSP isochrones we do not take into account
the effect of the unresolved  binaries  or stellar rotation   but focus on the possibility that any
unusual broadness at the MSTO might come from the presence of populations of different ages. 
Mackey et al. \shortcite{metal08},  Milone et al. \shortcite{metal09},
 Goudfrooij et al. \shortcite{getal09}, Piatti \shortcite{p13}, among others, showed that
a significant fraction of unresolved binaries is not enough to reproduce 
the EMSTOs seen in their studied clusters, while stellar rotation has not driven the
whole MSTO broadness in all the cases \cite{gietal09,letal14}. Since the purpose of this work
consists in introducing the high-quality deep photometric data set and provide with
mean cluster ages from which we will study any possible extended MSTO cluster candidate, 
the matching of SSP isochrones results overall justified. Moreover, by closely inspecting the
matched cluster MSTO regions we have a hint for any uncommon broadness
in the studied cluster sample.

In the matching procedure with a naked eye, we used seven different isochrones, ranging 
from slightly younger than the derived cluster age to slightly older. Finally, 
we adopted the cluster age as the age of the isochrone which best reproduced the cluster's 
main features in the CMD (namely, the cluster's MS, RCs and/or RGBs ). We noted, however, that the 
theoretically computed bluest stage during the He-burning core phase 
is redder than the observed RC in the CMDs of some clusters, a behaviour already 
detected in other studies of Galactic and Magellanic Cloud clusters (e.g., Piatti et al. 2009, Piatti et al. 
2011a,b, and references therein). A similar outcome was found from the  matching of isochrones in the 
$M_V$ vs $(V-I)_o$ plane \cite[among others]{petal03a,petal03b}. Figs. 16 to 19  show the results 
of isochrone matching. 

For each cluster CMD,  we plotted the isochrone of the adopted cluster age 
and two additional isochrones bracketing the derived age and separated by $|$$\Delta$(log($t$))$|$ = 0.05.  
The ages of the adopted isochrones for the cluster sample
are listed in Table 16. For clusters with density ratio ($\rho$ = cluster star density to background 
field star density ratio at r$_{HM}$) greater than 1.0, the age uncertainty is estimated as 0.05 dex in log($t$) while
 for clusters with larger background contamination, the uncertainty reaches 0.10 dex.
These age uncertainties are thought to mainly represent the overall dispersion along the SGB, RGB as well as 
the position of the RC, rather than a measure of the MSTO spread.
Nevertheless, in most of the clusters the adopted age uncertainties relatively reflect the observed MSTO broadness, 
thus implying a weaker chance for the EMSTO phenomenon. Note, however, that
we have assigned the same mean age error to NGC\,2173 and NGC\,2209 -even though only the latter was
confirmed as an EMSTO cluster by Keller et al. \cite{ketal12}-, simply because both clusters have
$\rho$ greater than 1.0.
In the last two columns we have compiled previously published age information. 
Fig. 20 shows the comparison between the published ages and
our present values. 
 The error bars correspond to the age uncertainties quoted by the authors,  while
the thick and thin lines represent the identity relationship and those shifted by 
$\pm$0.05, respectively. Black filled squares represent clusters that
do not fullfill the requirement 0.05 + $\sigma$(log($t$)$_{\rm pub}$)
$\ge$ $|$log($t$)$_{\rm our}$ - log($t$)$_{\rm pub}$ $|$. As can be seen, there is a reasonable
agreement ($|$present - literature values$|$ = 0.12 $\pm$ 0.10), although three clusters significantly 
depart from the $\pm$1$\sigma$ strip. In the case of NGC\,2161 and SL\,244, we checked that the yourger
ages by Geisler et al. \shortcite{getal03} and Piatti et al. \shortcite{petal11a}, respectively, are 
related to a much less deep photometry which barely reach the cluster MSTOs. In this sense, our
age estimates surpass in accuracy those previously derived. 
On the other hand, our age determination for NGC\,2249 
is significantly older than that determined by Baumgardt et al. \shortcite{betal13} but in line with the
 stated uncertainties of that work.
Finally, by looking at Figure 2 of Keller et al. \shortcite{ketal12}, the mean age of NGC\,2209 is 
log($t$) $\approx$ 9.06 dex, whereas we estimate a slightly older value (9.15). For NGC\,2173 the agreement 
is better: log($t$) $\approx$ 9.22 from their Figure 2 and 9.25 from our Table 16.

\section{Summary}

As part of a continuing project to investigate the extended MSTO phenomenon that is seen to be
 widespread in intermediate-age LMC clusters, we have used  the Gemini South telescope to obtain GMOS imaging in 
the SDSS $g,i$ system of 14 candidate intermediate-age LMC star clusters. Our aim is to establish 
a luminosity-limited sample of clusters in the age range of 1-3 Gyr in which to characterise the prevalence of the extended
MSTO phenomenon. In this work we present the CMDs of NGC\,2155, 2161, 
2162, 2173, 2203, 2209, 2213, 2231, 2249, Hodge\,6, SL\,244, 505, 674, and 769. The analysis of their photometric 
data leads to the following main conclusions:

(i) After extensive artificial star tests over the image data set, we show that the 50$\%$ 
completeness level is reached at $g,i$ $\sim$ 23.5-25.0, depending on the distance to the cluster centre, 
and that the behaviour of the photometric errors with magnitude for the observed stars guarantees the accuracy
of the morphology and position of the main features in the CMDs that we investigate.

(ii) We trace their stellar density radial profiles from star counts performed over the GMOS field of view.
From the density profiles, we adopted cluster radii defined as the distance from the cluster centre 
where the stellar density profile intersects the background level, and derived the radii at the full width at
half maximum of the radial profile.  We then built CMDs with cluster features clearly identified.

(iii) Using the cluster ($g,g-i$) diagrams, we estimated ages from theoretical 
isochrones computed for the SDSS system. The studied LMC clusters are confirmed to be intermediate-age
clusters of age, log($t$) = 9.10-9.60; we identified two of them, namely NGC 2162 and 2249, to be new 
extended MSTO cluster candidates on the basis of their dual red clumps.

\section*{Acknowledgements}
This work was partially supported by the Argentinian institutions CONICET and
Agencia Nacional de Promoci\'on Cient\'{\i}fica y Tecnol\'ogica (ANPCyT). 
SCK and GDC acknowledge the support of Australian Research Council (ARC) Discovery Project
grant DP120101237. ADM is grateful for support from an ARC Australian Research Fellowship (DP1093431). Based
 on observations obtained at the Gemini Observatory (Programs: GS-2011A-Q-43, GS-2012A-Q-15, and 
GS-2013A-Q-17), which is operated by the
Association of Universities for Research in Astronomy, Inc., under a cooperative agreement
with the NSF on behalf of the Gemini partnership: the National Science Foundation (United
States), the Science and Technology Facilities Council (United Kingdom), the National Research Council (Canada), 
CONICYT (Chile), the Australian Research Council (Australia),
Minist\'erio da Ci\`encia, Tecnologia e Inova\c{c}\~ao (Brazil) and Ministerio de Ciencia, Tecnolog\'{\i}a e
Innovaci\'on Productiva (Argentina).  We thank the anonymous referee whose comments and suggestions
allowed us to improve the manuscript.

\newpage

\begin{flushleft}
\begin{table}
\caption{Observation log of selected clusters.}
\begin{tabular}{@{}lcccccc}\hline
Star Cluster  & $\alpha_{\rm 2000}$ & $\delta_{\rm 2000}$  & 
filter & exposures & airmass & seeing  \\
     & (h m s)  & ($\degr$ $\arcmin$ $\arcsec$) & 
     &    (times $\times$ sec) &         & ($\arcsec$)\\
\hline

NGC\,2155 & 05 58 33 & -65 28 37 &  $g$ & 4$\times$150 + 4$\times$30 & 1.24-1.25 & 0.42-0.52 \\
          &          &           & $i$ & 4$\times$150 + 4$\times$15 & 1.25-1.27 & 0.35-0.41 \\
NGC\,2161 & 05 55 42 & -74 21 14 &  $g$ & 4$\times$150 + 4$\times$30 &  1.38     & 0.46-0.65 \\
          &          &           &    $i$ & 4$\times$150 + 4$\times$15 &  1.38     & 0.38-0.43 \\
NGC\,2162 & 06 00 31 & -63 43 17 &  $g$ & 4$\times$150 + 4$\times$30 & 1.53-1.59 & 0.55-0.64 \\ 
          &          &           &    $i$ & 4$\times$150 + 4$\times$15 & 1.60-1.66 & 0.52-0.72 \\
NGC\,2173 & 05 57 58 & -72 58 43 &  $g$ & 4$\times$150 + 4$\times$30 & 1.36      & 0.44-0.49 \\
          &          &           & $i$ & 4$\times$150 + 4$\times$15 & 1.36-1.37 & 0.35-0.42 \\
NGC\,2203 & 06 04 42 & -75 26 16 &  $g$ & 4$\times$150 + 4$\times$30 & 1.46-1.48 & 0.55-0.59 \\
          &          &           &   $i$ & 4$\times$150 + 4$\times$15 & 1.48.1.50 & 0.48-0.59 \\
NGC\,2209 & 06 08 34 & -73 50 28 &  $g$ &10$\times$145 + 4$\times$30 & 1.47-1.57 & 0.59-0.75 \\ 
          &          &           &  $i$ & 6$\times$137 + 4$\times$15 & 1.52-1.58 & 0.43-0.58 \\
NGC\,2213 & 06 10 42 & -71 31 44 & $g$ & 4$\times$150 + 4$\times$30 & 1.43-1.46 & 0.55-0.58 \\
          &          &           &   $i$ & 4$\times$150 + 4$\times$15 & 1.46-1.49 & 0.43-0.45 \\
NGC\,2231 & 06 20 44 & -67 31 05 &  $g$ & 5$\times$150 + 8$\times$30 & 1.43-1.46 & 0.58-0.68 \\
          &          &           &  $i$ & 4$\times$150 + 5$\times$15 & 1.47-1.51 & 0.58-0.64 \\
NGC\,2249 & 06 25 49 & -68 55 12 &  $g$ &12$\times$300 +15$\times$40 & 1.27.1.43 & 0.52-0.68 \\ 
          &          &           &   $i$ & 5$\times$280 + 5$\times$40 & 1.28-1.29 & 0.42-0.48 \\
Hodge\,6  & 05 42 17 & -71 35 28 &  $g$ & 9$\times$300 + 8$\times$40 & 1.33-1.40 & 0.43-0.78 \\
          &          &           &   $i$ & 6$\times$280 + 6$\times$40 & 1.33-1.40 & 0.48-0.58 \\
SL\,244   & 05 07 37 & -68 32 30 & $g$ & 4$\times$60                 & 1.48-1.50 &  0.69-0.77 \\
          &          &           & $i$ & 4$\times$30                 & 1.40-1.41 &  0.48-0.55 \\
SL\,505   & 05 28 50 & -71 38 00 & $g$ & 4$\times$60                 & 1.45-1.46 &  0.63-0.71 \\
          &          &           & $i$ & 4$\times$30                 & 1.39-1.40 &  0.48-0.58 \\
SL\,674   & 05 43 20 & -66 15 42 & $g$ & 4$\times$60                 & 1.40-1.42 &  0.57-0.70 \\
          &          &           & $i$ & 4$\times$30                 & 1.32-1.33 &   0.47-0.53 \\
SL\,769   & 05 53 23 & -70 04 18 & $g$ & 4$\times$60                 & 1.40-1.41 &  0.63-0.64 \\
          &          &           & $i$ & 4$\times$30                 & 1.35      &   0.45-0.55 \\
\hline
\end{tabular}
\end{table}
\end{flushleft}

\begin{table}
\caption{CCD $gi$ data of stars in the field of NGC\,2155.}
\begin{tabular}{@{}lccccccc}\hline
Star & RA(J2000)  & DEC(J2000) & $g$ & $\sigma$($g$) & $g-i$ & $\sigma$$(g-i)$ & n \\
     & (h:m:s) & ($\deg$ $\arcmin$ $\arcsec$) & (mag) & (mag) & (mag) & (mag)  \\\hline
-    &   -     &   -     &  -    &  -    &   -   &   -     \\
      9 & 05:58:13.149 & -65:31:27.72 &  22.835  &  0.014  &  0.275  &  0.022 &  4 \\
     10 & 05:58:09.841 & -65:30:31.02 &  24.139  &  0.023  &  0.347  &  0.067 &  4 \\
     11 & 05:58:03.581 & -65:28:43.59 &  24.139  &  0.038  &  0.465  &  0.053 &  3 \\
-    &   -     &   -     &  -    &  -    &   -   &   -     \\
\hline
\end{tabular}
\end{table}

\setcounter{table}{15}
\begin{table}
\caption{Fundamental properties of LMC star clusters.}
\begin{tabular}{@{}lccccccc}\hline
Star Cluster & r$_{HM}$ & r$_{cls}$ & $\rho$ & $E(B-V)$ &  log($t$) & log($t$) & Ref. \\
             &   (pc) &  (pc) & &(mag) &          &   (literature) & \\
\hline
NGC\,2155   & 8.5 & 26.5 & 3.1 & 0.03  &   9.50 $\pm$ 0.05 &   9.56 $\pm$ 0.08 & 2\\
NGC\,2161   & 5.8 & 26.5 & 2.8 &0.12  &   9.35 $\pm$  0.05 &   9.04 $\pm$ 0.12 & 3\\
NGC\,2162   & 7.8 & 24.8 & 4.5 &0.03  &   9.20 $\pm$  0.05  & 9.11 $\pm$ 0.14 &  1 \\
NGC\,2173   & 9.8 & 28.3 & 1.7 &0.09  &   9.25 $\pm$  0.05 &  9.33 $\pm$ 0.08 &  1 \\
NGC\,2203   & 9.5 & 31.8 & 8.1 &0.11  &   9.30 $\pm$  0.05&   &  \\
NGC\,2209   & 8.5 & 33.6 & 4.2 &0.11  &   9.15 $\pm$  0.05 &   9.18 $\pm$ 0.09 &  4 \\
NGC\,2213   & 5.7 & 21.2 & 3.7 &0.11  &   9.25 $\pm$  0.05  & 9.20 $\pm$ 0.11 &  1 \\
NGC\,2231   & 6.3 & 23.0 & 4.0 &0.06  &   9.20 $\pm$  0.05 & 9.18 $\pm$ 0.11 &  1 \\
NGC\,2249   & 6.8 & 21.2 & 4.1 &0.07  &   9.15 $\pm$  0.05 &   8.82 $\pm$ 0.30 &  1 \\
Hodge\,6    & 10.1 & 17.7 & 0.6 &0.09  &   9.40 $\pm$  0.10 &   &  \\
SL\,244     & 5.5 & 12.4 & 0.3 &0.07  &   9.40 $\pm$ 0.10  &   9.18 $\pm$ 0.09 & 5 \\
SL\,505     & 2.5 & 12.4 & 0.6 &0.08  &   9.30 $\pm$ 0.10  &  9.18 $\pm$ 0.09 & 5 \\
SL\,674     & 6.6 & 17.7 & 1.3 &0.05  &   9.45 $\pm$ 0.05  &   9.36 $\pm$ 0.06 & 5 \\
SL\,769     & 7.0 & 17.7 & 0.9 &0.08  &  9.35 $\pm$ 0.10  &   9.25 $\pm$ 0.07 & 6 \\

\hline
\end{tabular}
\medskip

\noindent Ref.: 1) Baumgardt et al. 2013; 2) Piatti et al. 2002; 3) Piatti et al. 2011a; 4) 
Piatti et al. 1999; 5) Geisler et al 2003; 6) Bica et al. 1998.

\end{table}

\clearpage

\begin{figure}
\includegraphics[width=167mm]{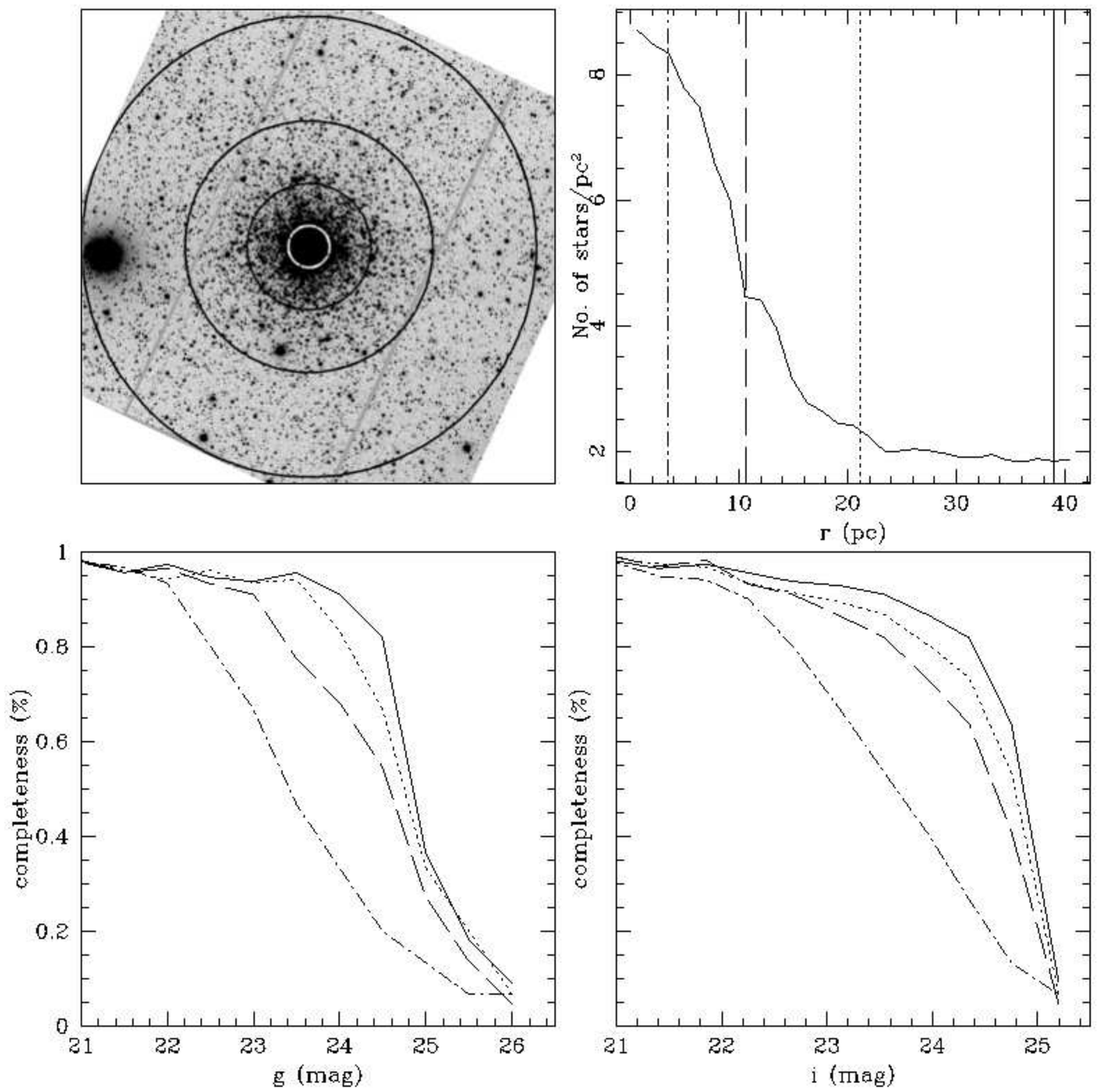}
\caption{$i$ image with the LMC cluster NGC\,2173 and four circles of 100, 300, 600, and 1100 pixels overplotted (upper left). 
North is up and east is to the left. The cluster density profile (upper right) and the completeness level in
$g$ (bottom left) and $i$ (bottom-right) bands are also shown for different circular rings: 0-100 pixels 
(dot dashed line); 100-300 pixels (dashed line); 300-600 pixels (dotted line), and 600-1100 pixels (solid line).}
\label{fig1}
\end{figure}

\begin{figure}
\includegraphics[width=167mm]{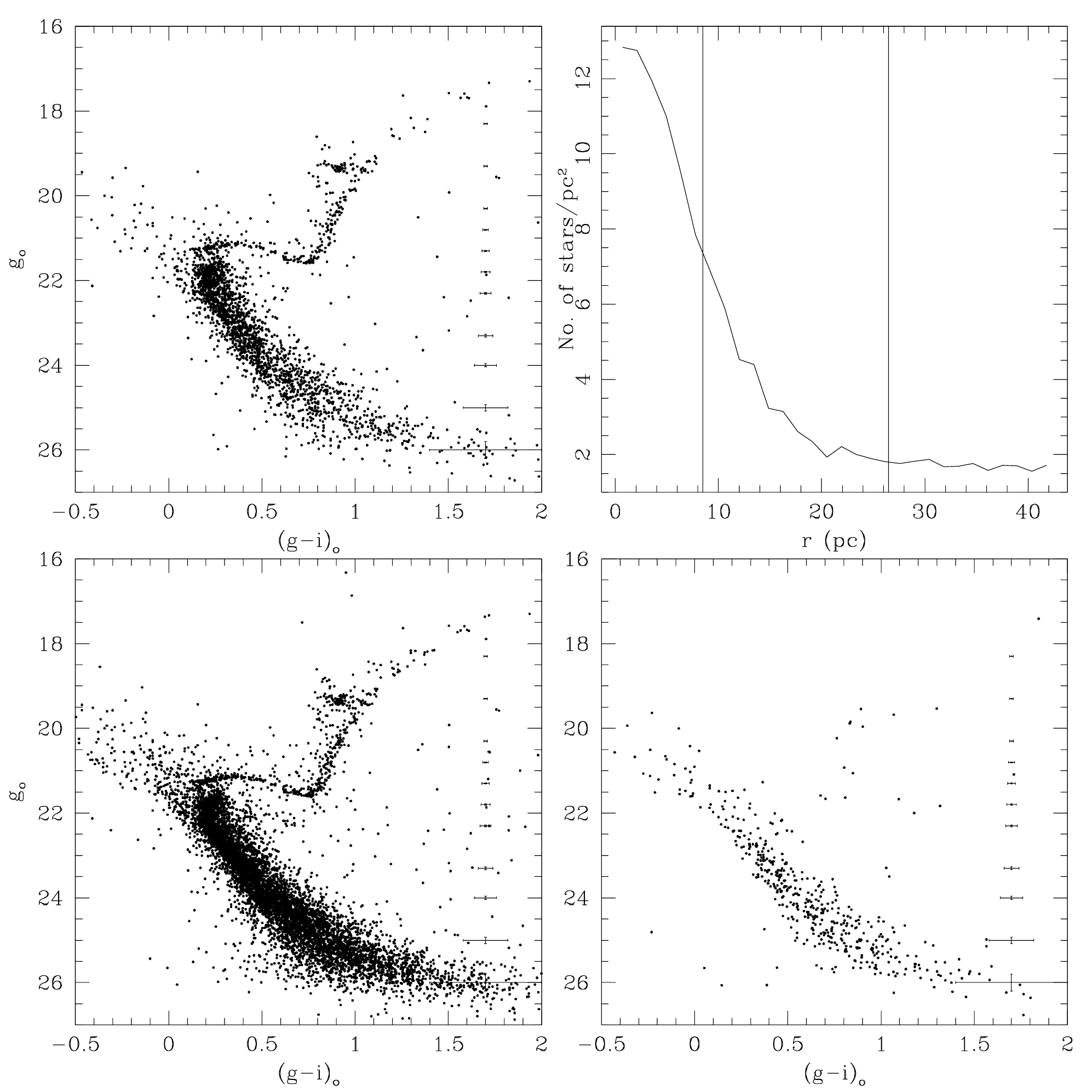}
\caption{Colour-magnitude diagrams for stars in NGC\,2155 distributed within circles centred on the cluster
and radii $r_{HM}$ (upper left) and $r_{cls}$ (bottom left), and that
for surrounding field stars distributed within a ring of area $\pi$$r_ {HM}$$^2$ (bottom right).
The cluster density profile with the radii at $r_{HM}$ and $r_{cls}$ indicated  is also shown (uppe right).}
\label{fig2}
\end{figure}

\begin{figure}
\includegraphics[width=167mm]{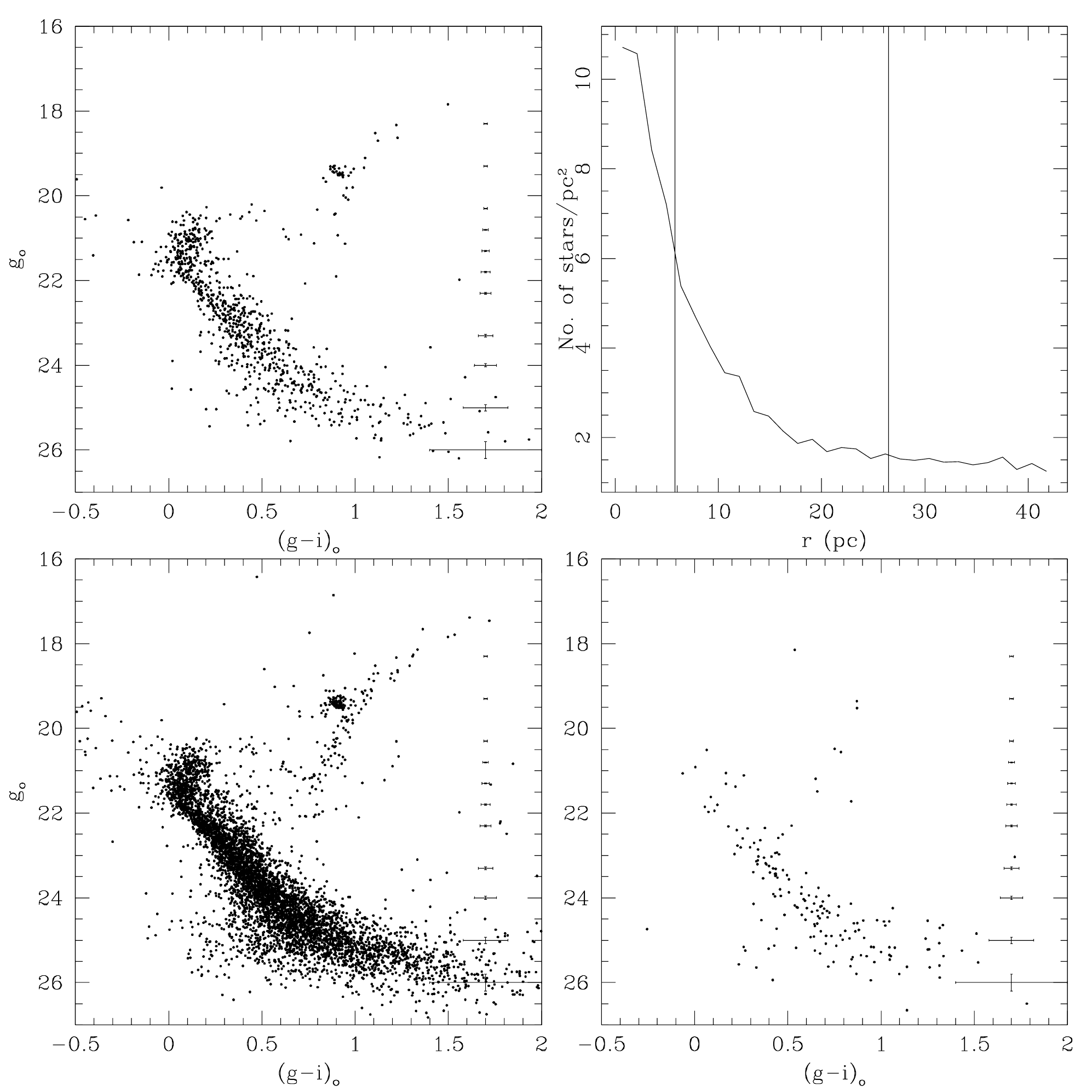}
\caption{Idem as Fig. 2 for NGC\,2161.}
\label{fig3}
\end{figure}

\begin{figure}
\includegraphics[width=167mm]{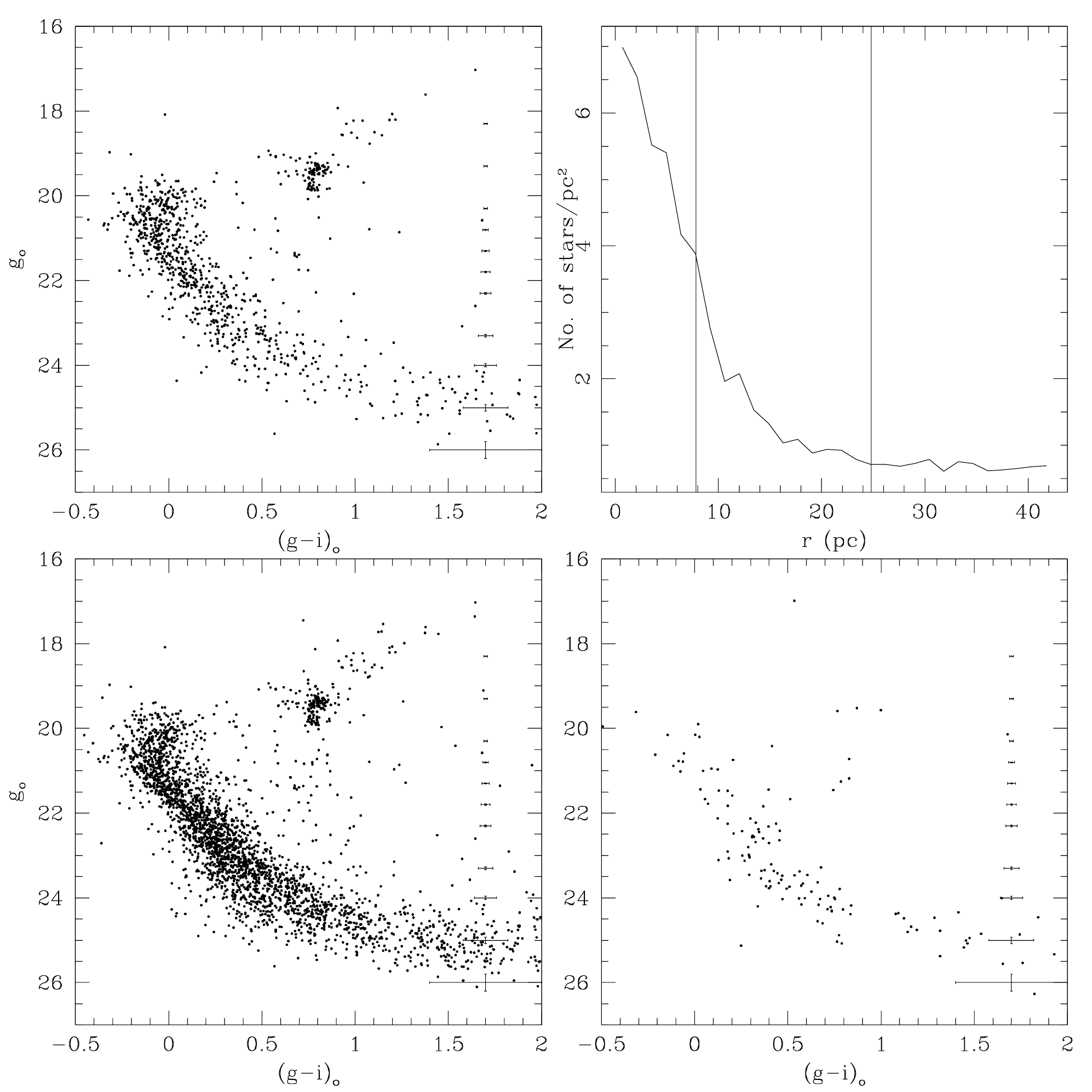}
\caption{Idem as Fig. 2 for NGC\,2162.}
\label{fig4}
\end{figure}

\begin{figure}
\includegraphics[width=167mm]{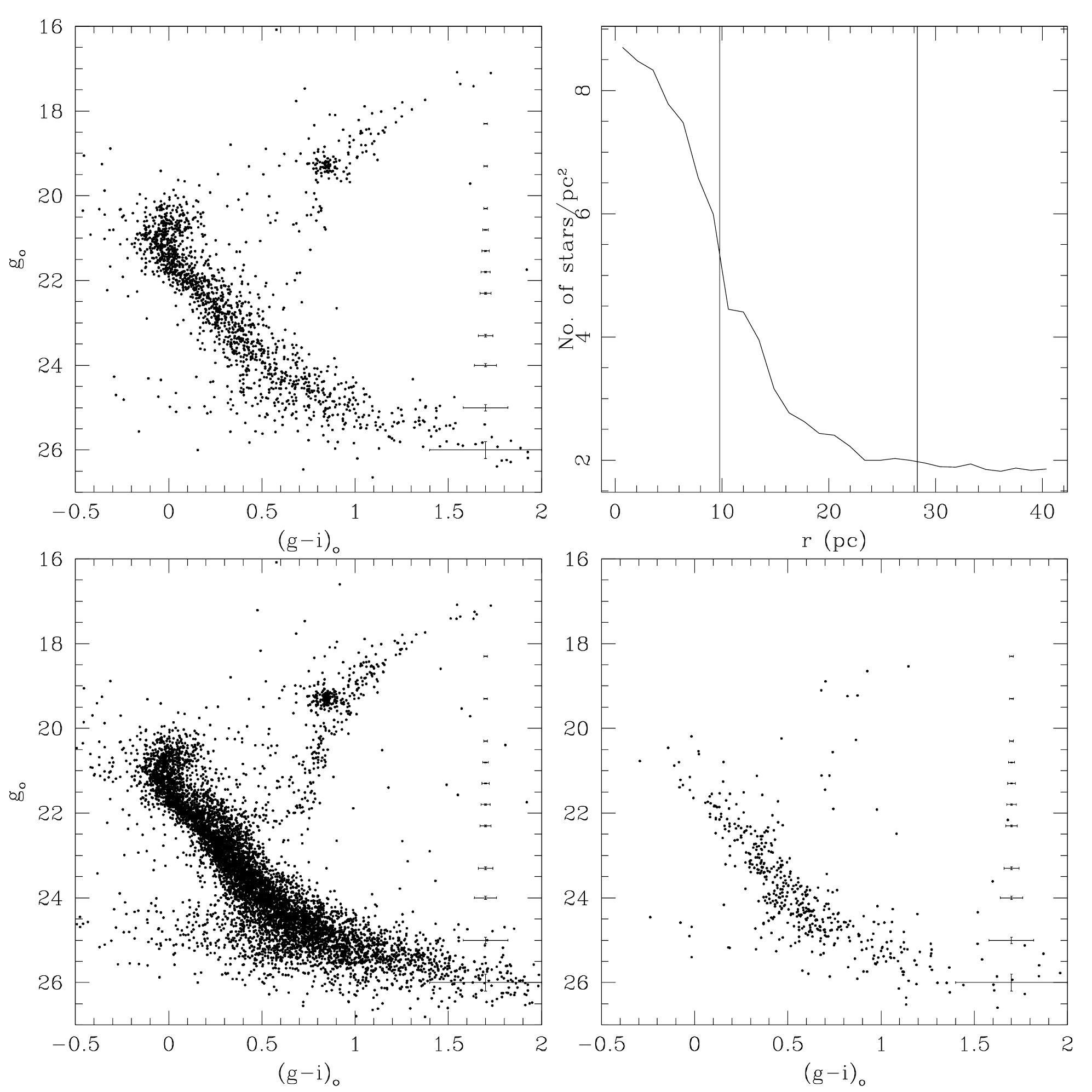}
\caption{Idem as Fig. 2 for NGC\,2173.}
\label{fig5}
\end{figure}

\begin{figure}
\includegraphics[width=167mm]{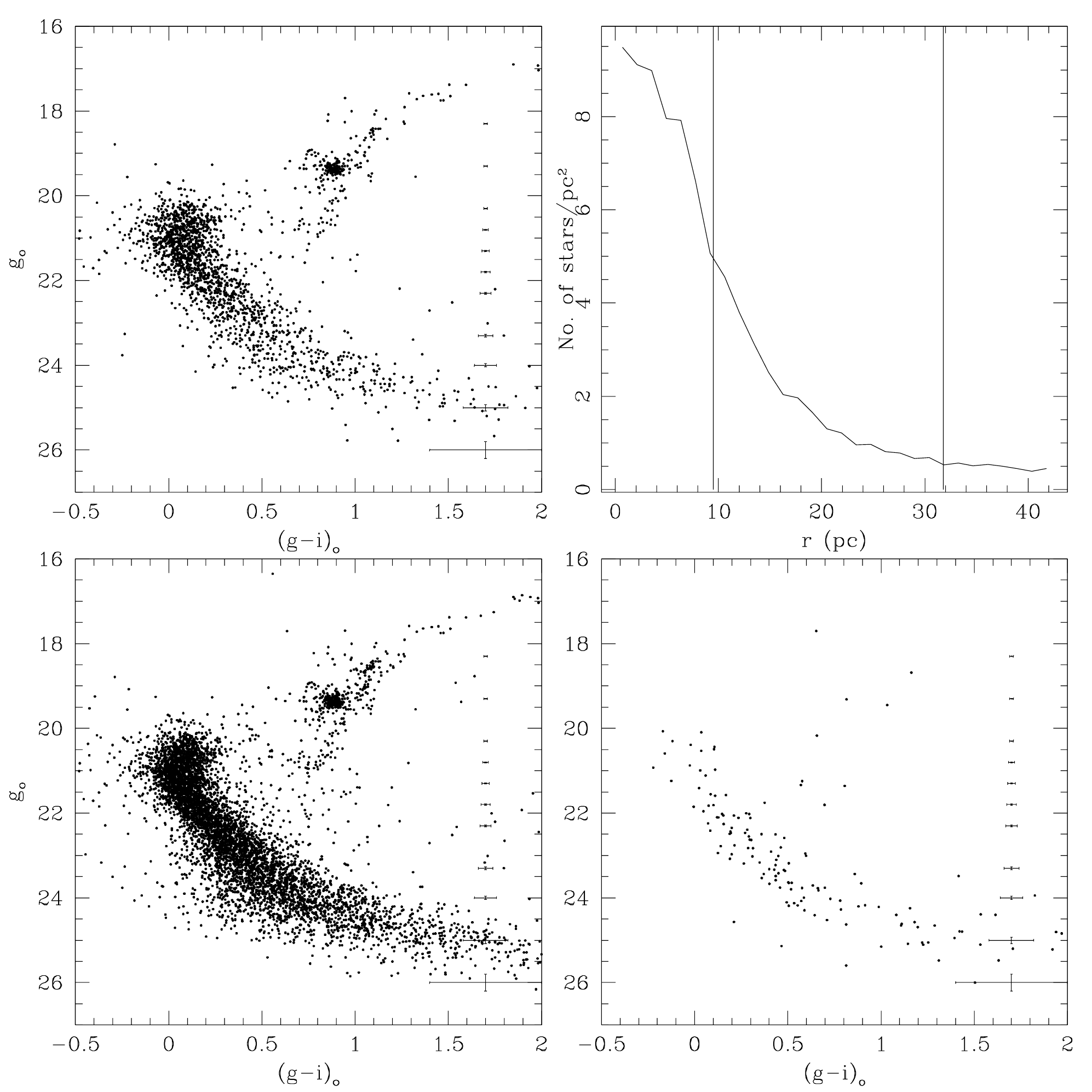}
\caption{Idem as Fig. 2 for NGC\,2203.}
\label{fig6}
\end{figure}

\begin{figure}
\includegraphics[width=167mm]{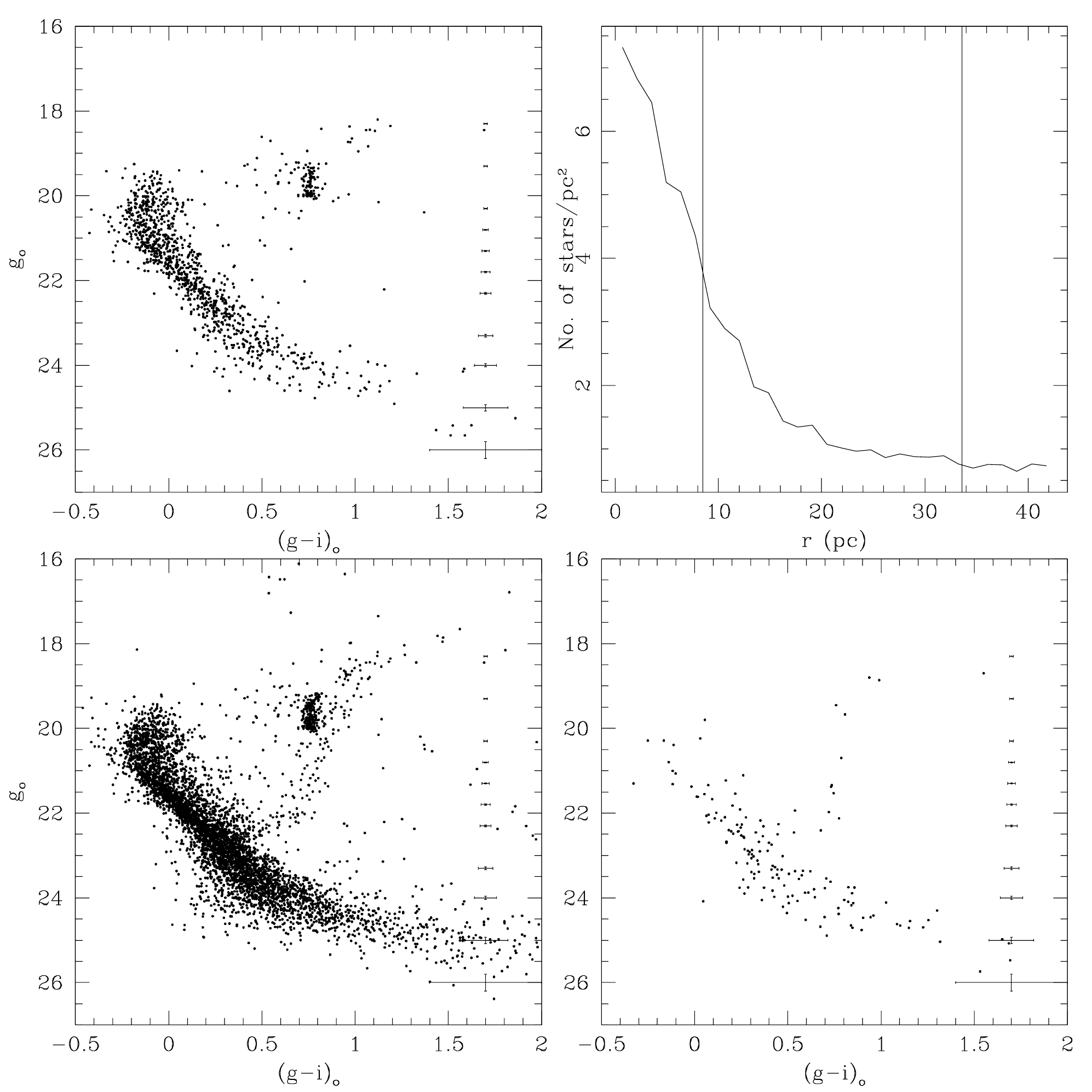}
\caption{Idem as Fig. 2 for NGC\,2209.}
\label{fig7}
\end{figure}

\begin{figure}
\includegraphics[width=167mm]{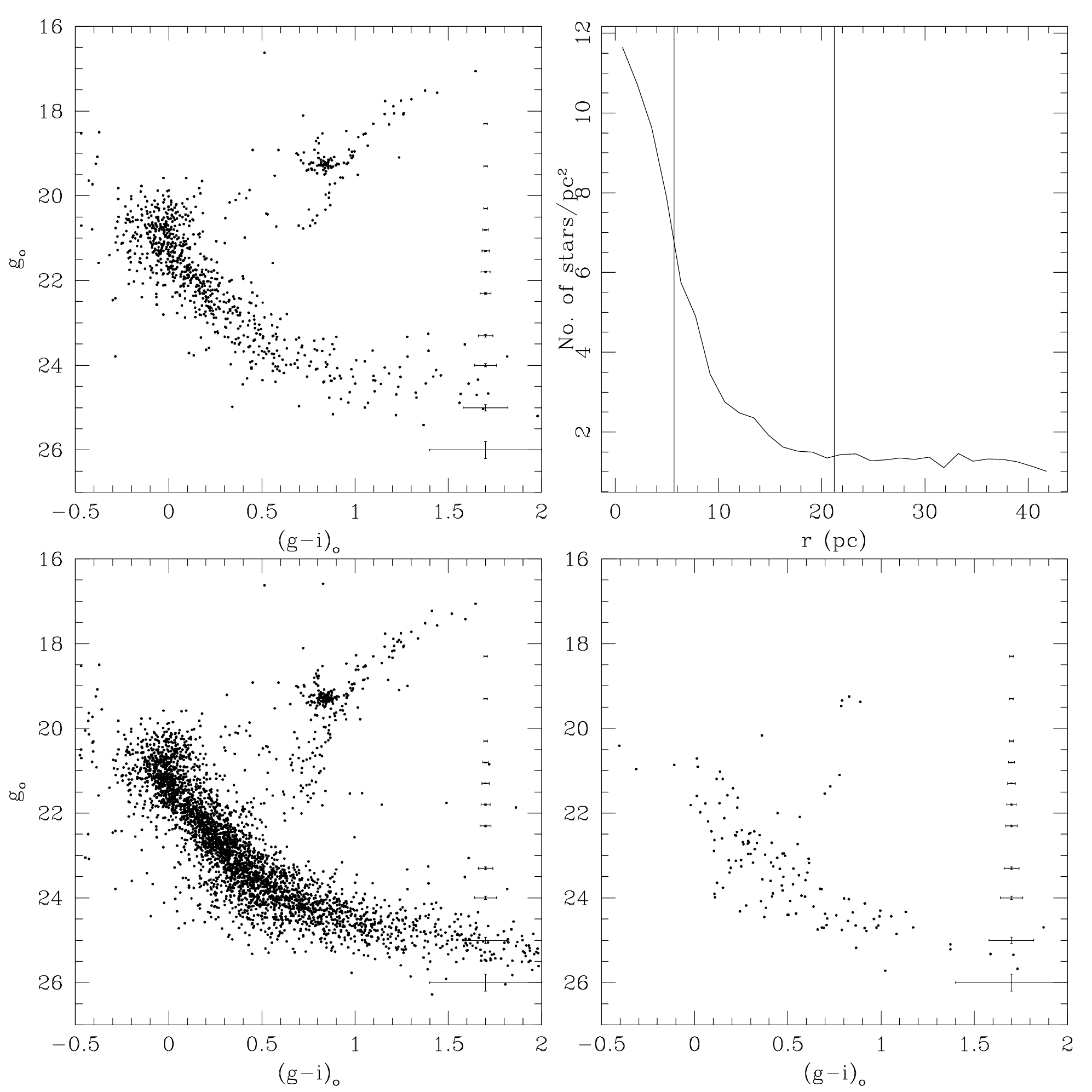}
\caption{Idem as Fig. 2 for NGC\,2213.}
\label{fig8}
\end{figure}

\begin{figure}
\includegraphics[width=167mm]{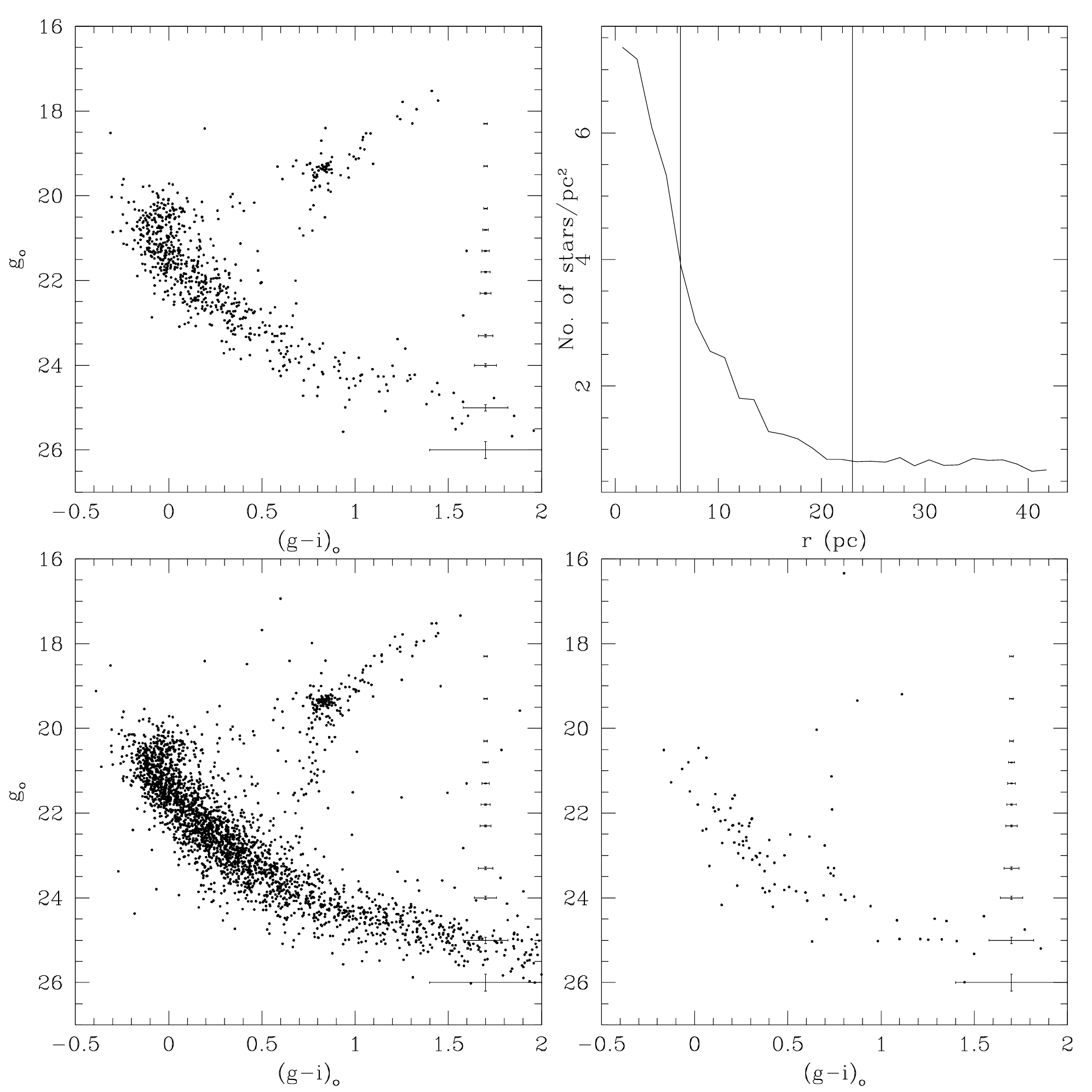}
\caption{Idem as Fig. 2 for NGC\,2231.}
\label{fig9}
\end{figure}

\begin{figure}
\includegraphics[width=167mm]{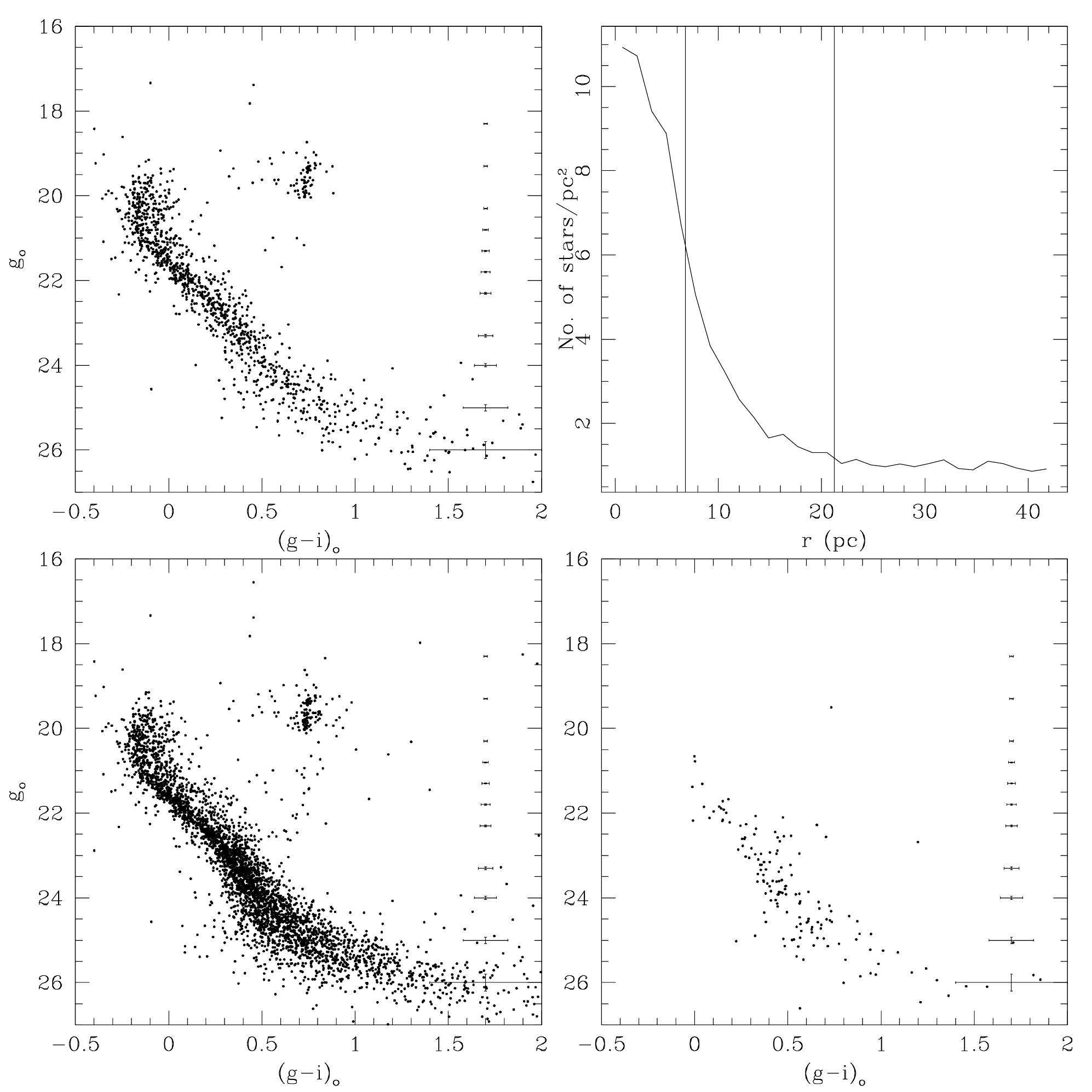}
\caption{Idem as Fig. 2 for NGC\,2249.}
\label{fig10}
\end{figure}

\clearpage

\begin{figure}
\includegraphics[width=167mm]{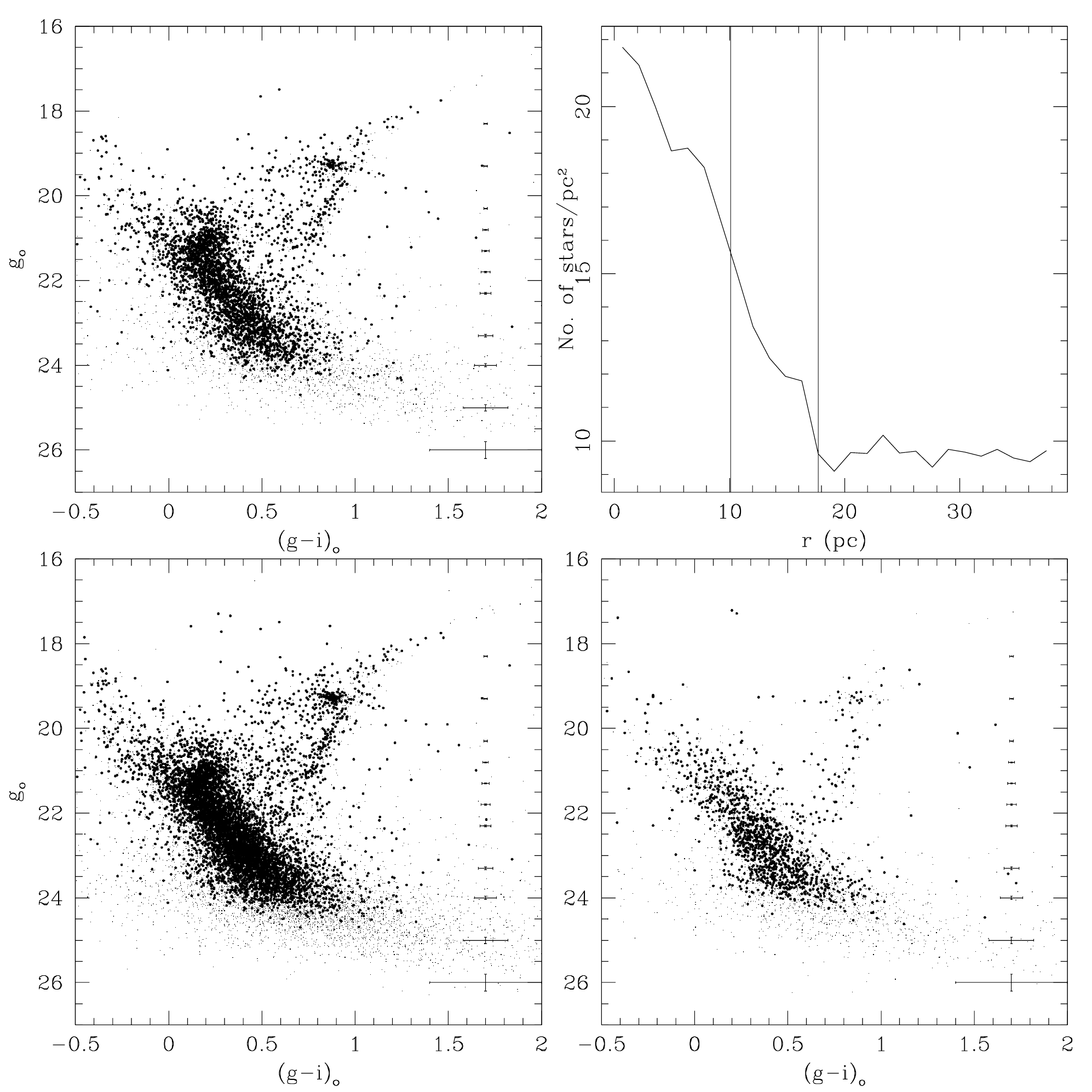}
\caption{Idem as Fig. 2 for Hodge\,6. Small points correspond to stars with a number of observations less than
5.}
\label{fig11}
\end{figure}

\begin{figure}
\includegraphics[width=167mm]{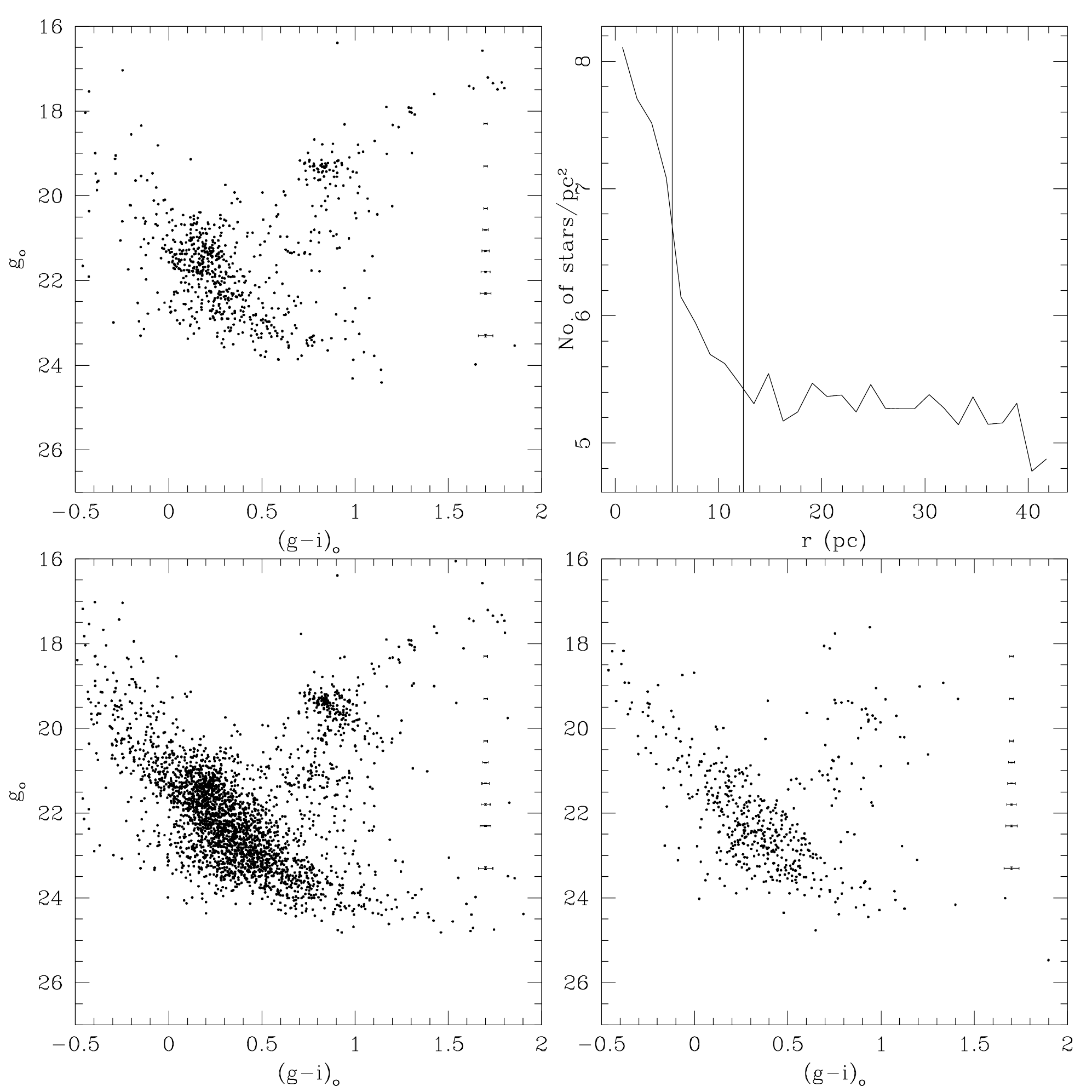}
\caption{Idem as Fig. 2 for SL\,244.}
\label{fig12}
\end{figure}

\begin{figure}
\includegraphics[width=167mm]{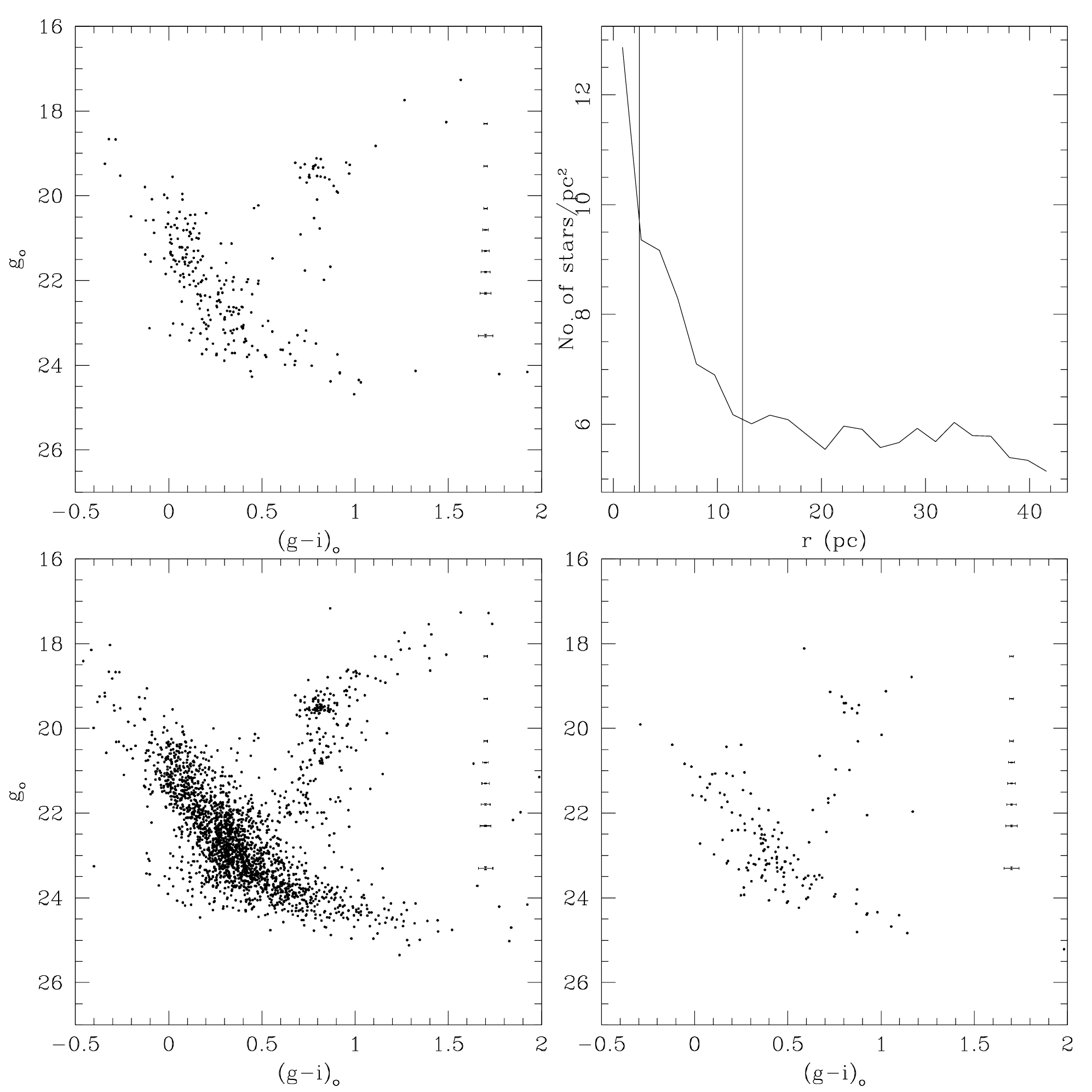}
\caption{Idem as Fig. 2 for SL\,505.}
\label{fig13}
\end{figure}

\begin{figure}
\includegraphics[width=167mm]{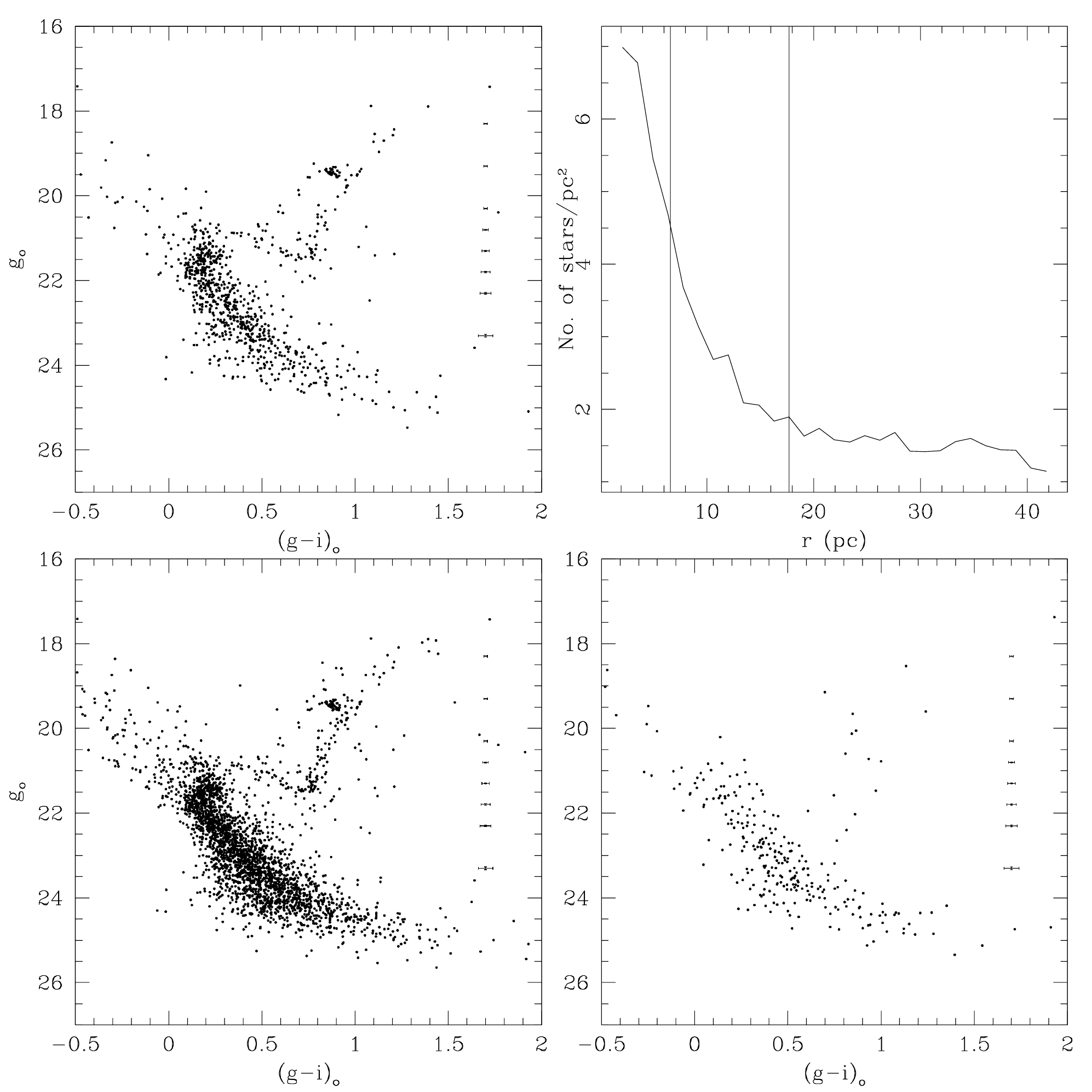}
\caption{Idem as Fig. 2 for SL\,674.}
\label{fig14}
\end{figure}

\begin{figure}
\includegraphics[width=167mm]{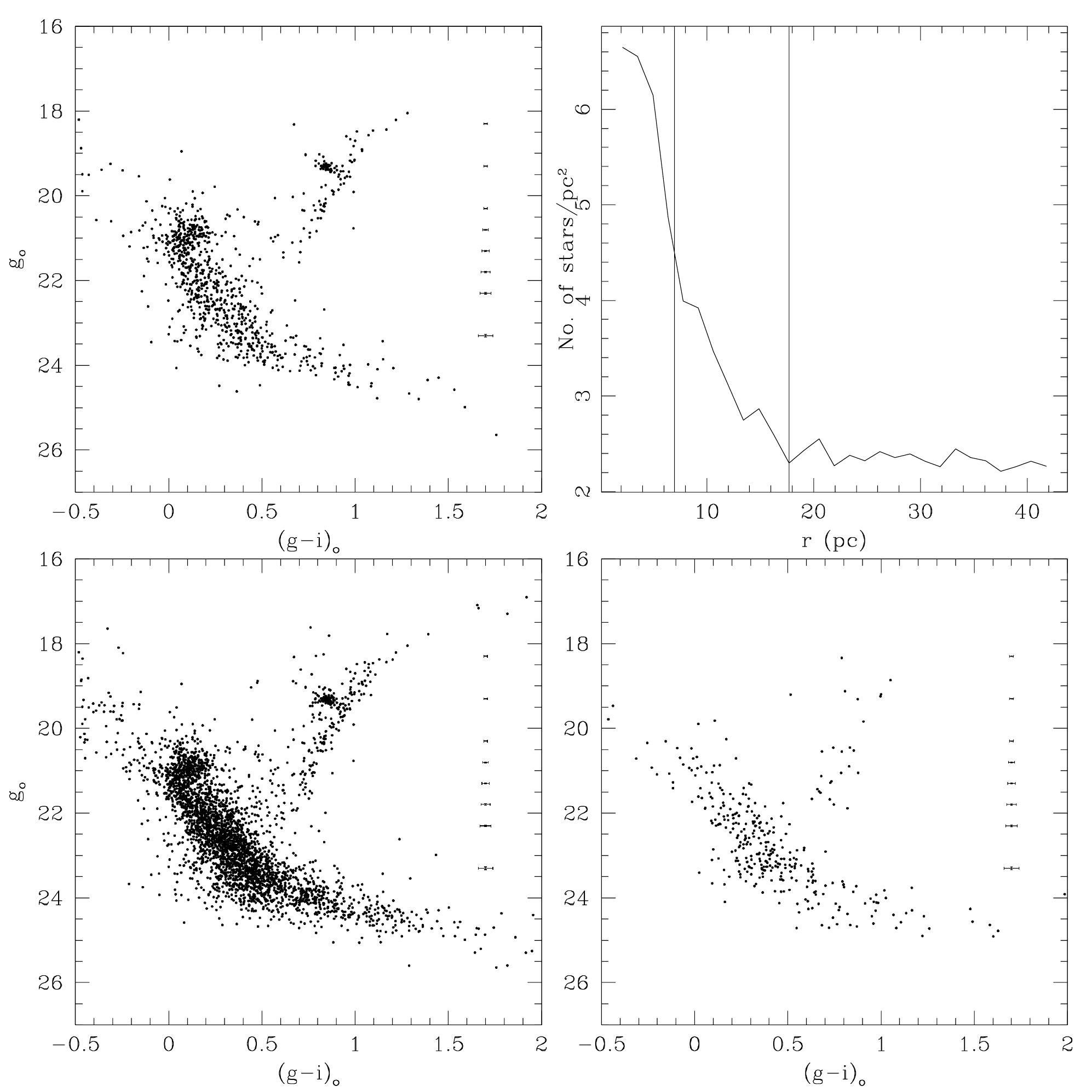}
\caption{Idem as Fig. 2 for SL\,769.}
\label{fig15}
\end{figure}

\begin{figure}
\includegraphics[width=167mm]{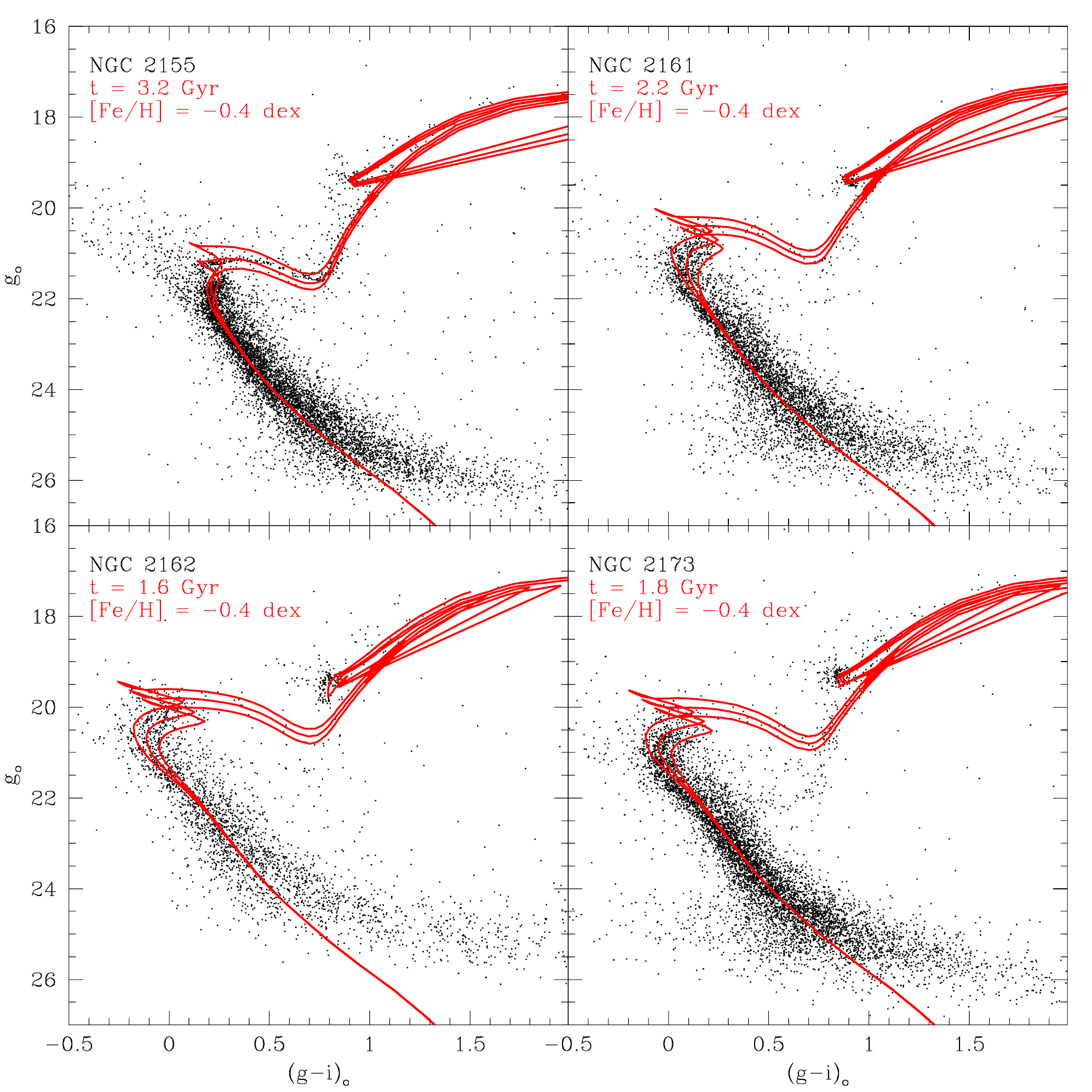}
\caption{Theoretical isochrones from Marigo et al. (2008) superimposed on to LMC
cluster CMDs. The youngest isochrone correspondes to log($t$) - $\sigma$(log($t$)) and
metallicity (Z) listed in Table 16, whereas the isochrone separation is $\Delta$(log($t$)) = 0.05.}
\label{fig16}
\end{figure}

\begin{figure}
\includegraphics[width=167mm]{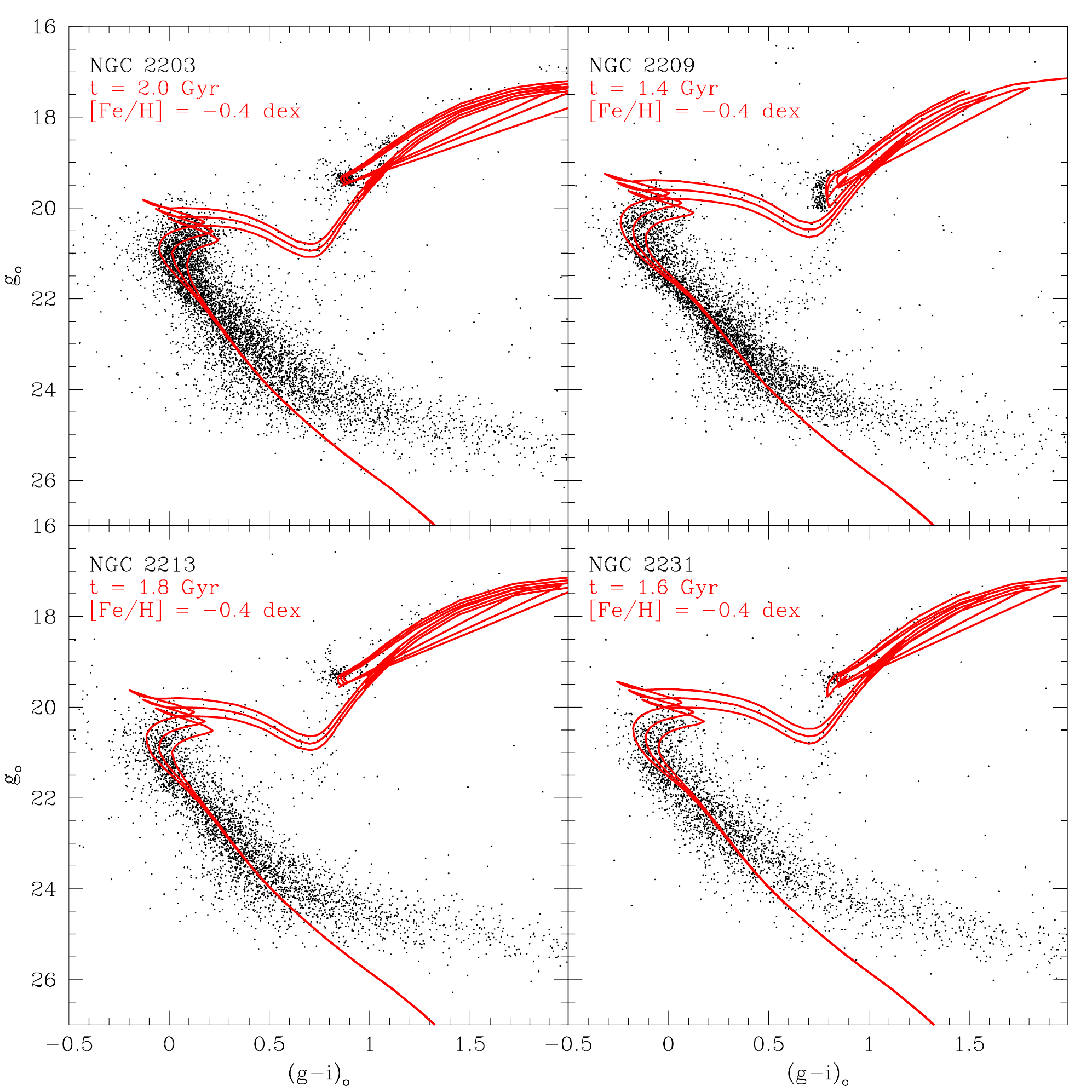}
\caption{Theoretical isochrones from Marigo et al. (2008) superimposed on to LMC
cluster CMDs. The youngest isochrone correspondes to log($t$) - $\sigma$(log($t$)) and
metallicity (Z) listed in Table 16, whereas the isochrone separation is $\Delta$(log($t$)) = 0.05.}
\label{fig17}
\end{figure}

\begin{figure}
\includegraphics[width=167mm]{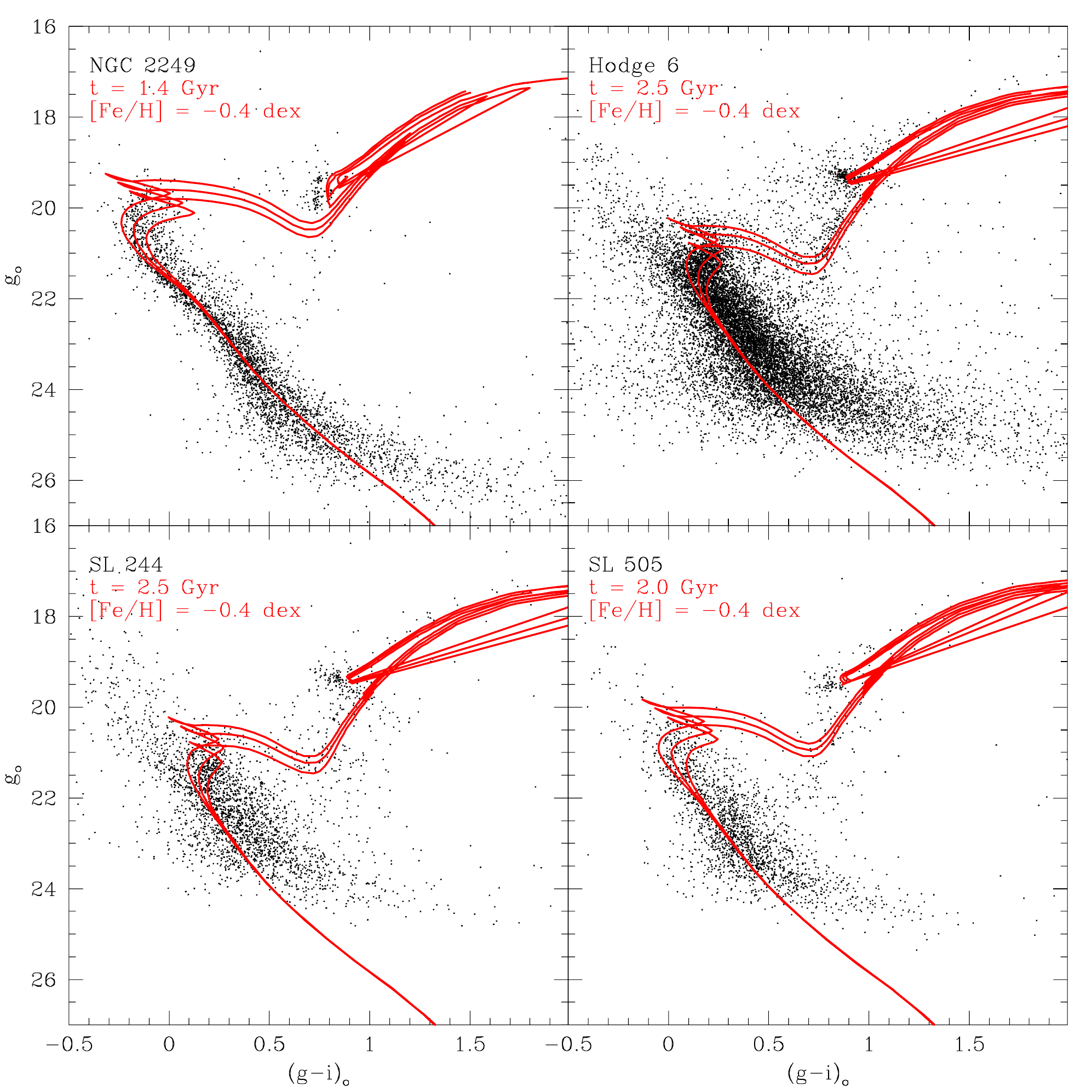}
\caption{Theoretical isochrones from Marigo et al. (2008) superimposed on to LMC
cluster CMDs. The youngest isochrone correspondes to log($t$) - $\sigma$(log($t$)) and
metallicity (Z) listed in Table 16, whereas the isochrone separation is $\Delta$(log($t$)) = 0.05.}
\label{fig18}
\end{figure}

\begin{figure}
\includegraphics[width=167mm]{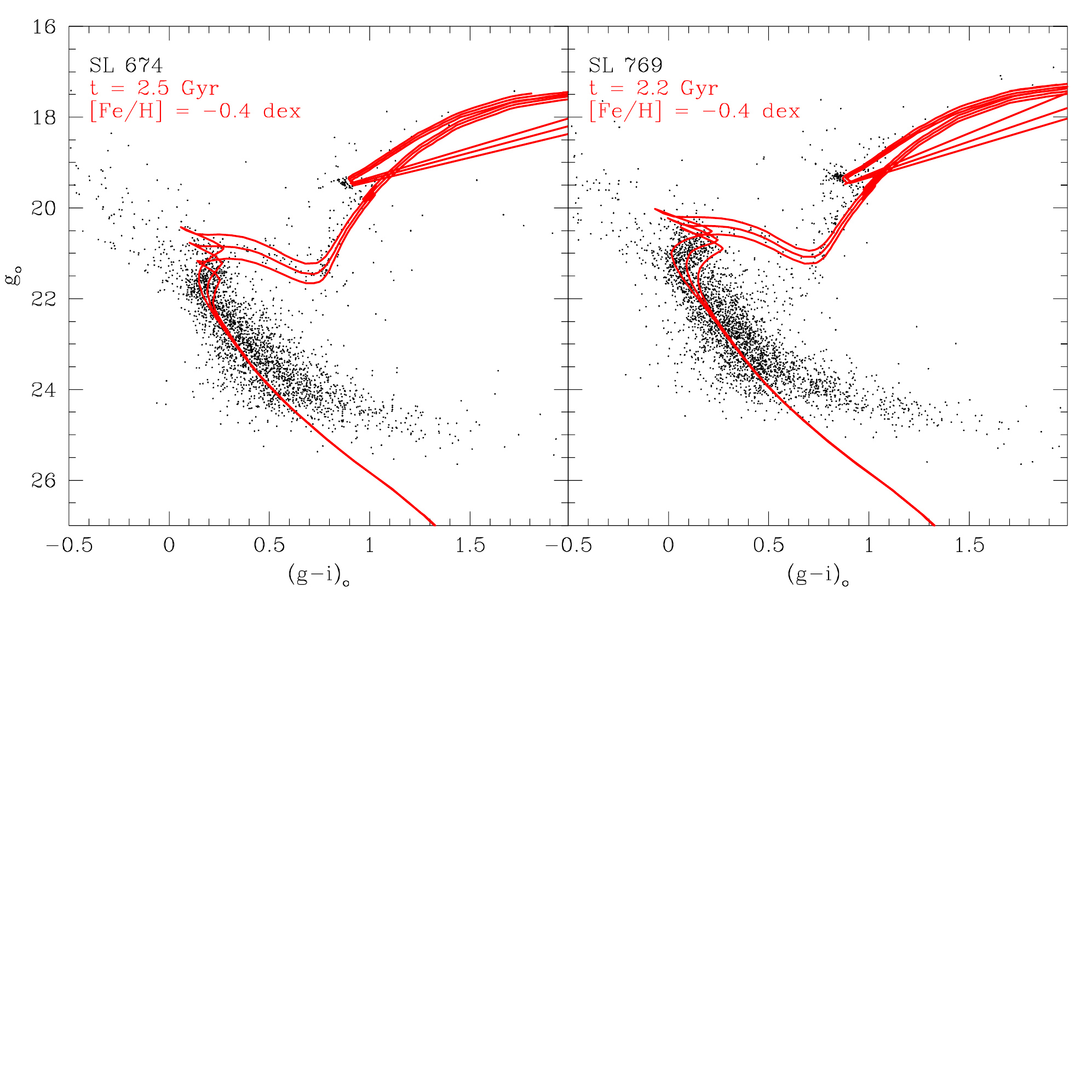}
\caption{Theoretical isochrones from Marigo et al. (2008) superimposed on to LMC
cluster CMDs. The youngest isochrone correspondes to log($t$) - $\sigma$(log($t$)) and
metallicity (Z) listed in Table 16, whereas the isochrone separation is $\Delta$(log($t$)) = 0.05.}
\label{fig19}
\end{figure}

\begin{figure}
\includegraphics[width=167mm]{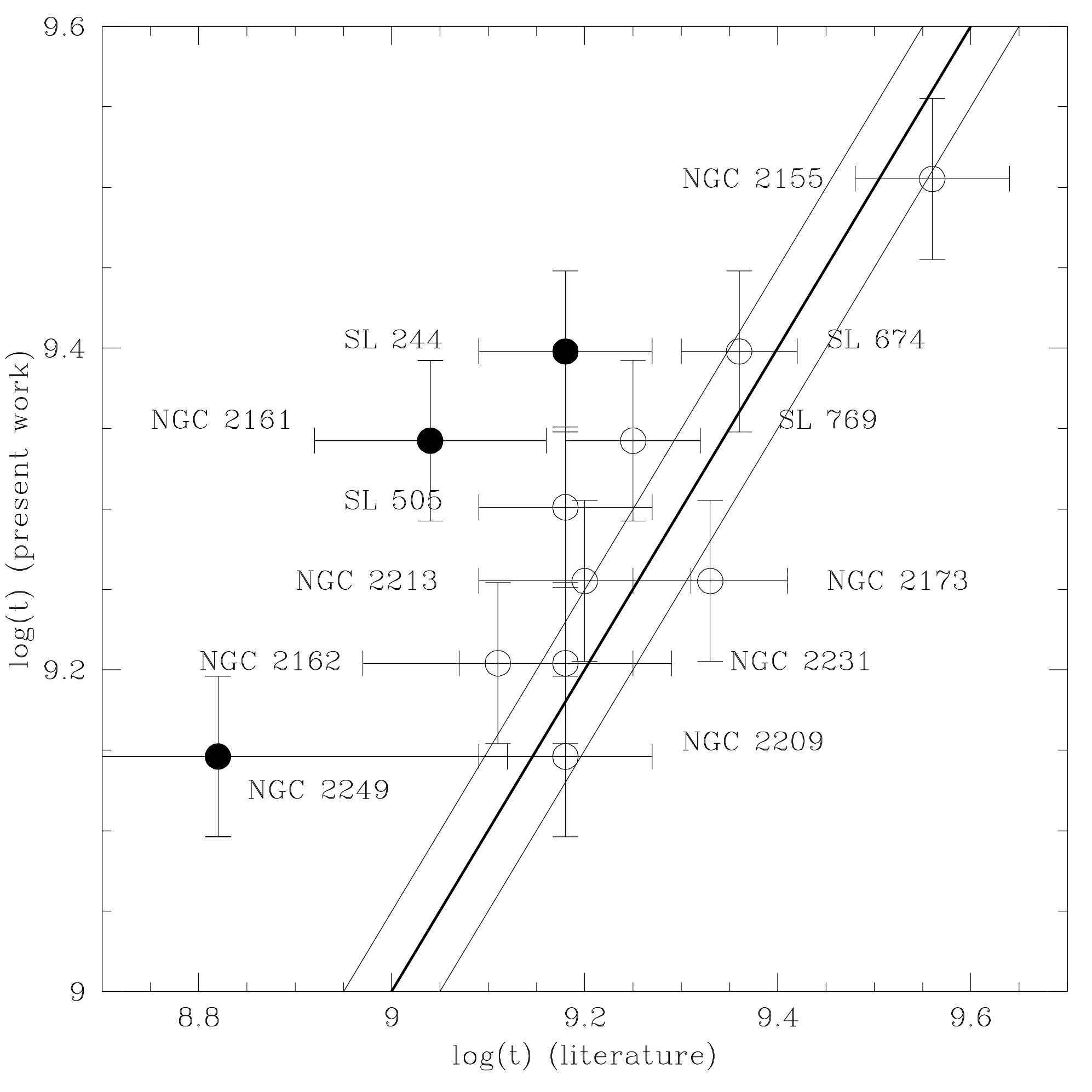}
\caption{Comparison of LMC cluster ages determined in this work with those
available in the literature. The straight line represents the 1:1 relationship.}
\label{fig20}
\end{figure}

\end{document}